\documentclass{aa}
\usepackage{txfonts}
\usepackage{graphicx}
\usepackage{caption}
\usepackage{subcaption}
\usepackage{color}

\usepackage{tikz}
\usetikzlibrary{arrows,shapes,backgrounds}

\begin{document} 
\titlerunning{}
\authorrunning{Mountrichas et al.}
\titlerunning{Application of X-CIGALE in the XMM-XXL field}

\title{X-ray flux in the SED modelling: An application of X-CIGALE in the XMM-XXL field}

\author{G. Mountrichas\inst{1}, V. Buat\inst{2,3}, G. Yang\inst{4,5}, M. Boquien\inst{6}, D. Burgarella\inst{2}, L. Ciesla\inst{2}}
          
    \institute {Instituto de Fisica de Cantabria (CSIC-Universidad de Cantabria), Avenida de los Castros, 39005 Santander, Spain
              \email{gmountrichas@gmail.com}
           \and
             Aix Marseille Univ, CNRS, CNES, LAM Marseille, France. 
                \email{ veronique.buat@lam.fr}  
              \and
                 Institut Universitaire de France (IUF)
                \and
                Department of Physics and Astronomy, Texas A\&M University, College Station, TX 77843-4242, USA 
               \and
               	George P. and Cynthia Woods Mitchell Institute for Fundamental Physics and Astronomy, Texas A\&M University, College Station, TX 77843-4242, USA 
               	\and
               	Centro de Astronom\'ia (CITEVA), Universidad de Antofagasta, Avenida Angamos 601, Antofagasta, Chile}

\abstract {X-CIGALE, built upon the spectral energy distribution (SED) code of CIGALE, implements important new features: the code accounts for obscuring material in the polars of the AGN and has the ability to fit X-ray fluxes. In this work, we use $\sim 2500$ spectroscopic, X-ray AGN from the XMM-XXL-North field and examine the improvements the new features bring in the SED modelling analysis. Based on our results, X-CIGALE successfully connects the X-ray with the UV luminosity in the whole range spanned by our sample ($\rm {log\,L_{X}(2-10\,keV)=(42-46)\, erg\,s^{-1}}$). The addition of the new features globally  improves the efficiency of X-CIGALE in the estimation and characterization of the AGN component. The classification into type 1 and type 2 based on their inclination angle is improved, especially at redshifts lower than 1. The statistical significance of the AGN fraction, $\rm frac_{AGN}$, measurements is increased, in particular for luminous X-ray sources ($\rm {L_{X}>10^{45}\, erg\,s^{-1}}$). These conclusions hold under the condition that (mid-) IR photometry is available in the SED fitting process. The addition of polar dust increases the AGN fraction and the efficiency of the SED decomposition to detect AGN among X-ray selected sources. X-CIGALE estimates a strong AGN ($\rm frac_{AGN}>0.3$) in more than $90\%$ of the infrared selected AGN and $75\%$ of X-ray detected AGN not selected by IR colour criteria. The latter drops to $\sim 50\%$ when polar dust is not included. The ability of X-CIGALE to include X-ray information in the SED fitting process can be instrumental in the optimal exploitation of the wealth of data that current (eROSITA) and future (ATHENA) missions will provide us.} 

   
\maketitle

\section{Introduction}

One of the major challenges in current astrophysical research is to understand the physical processes that operate on top of the dark matter distribution to produce luminous structures, such as stars and galaxies. These baryonic physics has a high level of complexity mainly because it includes electromagnetic interactions, for instance the interplay between cooling and heating of baryons (e.g. gas of dust particles) via radiative processes, or the impact of magnetic fields on charged particles. However, such processes operate on spatial scales many orders of magnitude smaller than what can be achieved by state-of-the-art cosmological simulations. Thus, assumptions have to be made on the amplitude and impact of the multitude of possible mechanisms that may affect the formation and evolution of galaxies, e.g. gas supply, gas cooling and heating, impact of stellar winds on interstellar medium. These physical processes that govern the birth and fate of galaxies are of extreme complexity. However, they are of great interest since galaxies play a pivotal role in the structure of the Universe and are unique tracers of its evolution.

One manifestation of baryonic physics is the formation of supermassive black holes (SMBHs) at the centre of galaxies. Tight correlations have been found between the mass of the SMBH and the properties of its bulge \citep[e.g.][]{Magorrian1998, Ferrarese2000}. When material is accreted onto these SMBHs it triggers them and the galaxy is called Active Galactic Nuclei (AGN). The energy released during the accretion process is also an important source of heating for both the interstellar \citep[e.g.][]{Morganti2017} and intergalactic medium \citep[e.g.][]{Kaastra2014}. As a result, it has been hypothesised that SMBHs and their activity plays an important role in galaxy evolution \citep[e.g.][]{Brandt_Alexander2015}. 

AGN emission can be observed at different wavelengths from X-rays to radio, as different physical mechanisms produce radiation at different wavelengths. In particular, X-ray emission is a trademark of AGN activity. This emission originates from photons produced by the accretion disk that are scattered by the hot corona and emit X-rays through inverse Compton scattering. This process, dominates the X-ray emission of the host galaxy and reflects the activity of the central SMBH. Thus, X-rays are often used as a proxy of AGN power \citep[e.g.][]{Lusso2012, Yang2019}. Important ongoing (e.g., eROSITA) and future (ATHENA) X-ray missions, will use the unique window that X-rays offer and provide us a wealth of data to study the tight connection of AGN with their host galaxies.

The multi-wavelength emission of galaxies can be studied by constructing and modelling their full spectral energy distribution (SED). This method allows to measure fundamental properties of galaxies, such as their stellar mass $\rm M_*$, star formation rate (SFR), dust mass and attenuation while at the same time breaks degeneracies that plague observations in narrow(er) wavelength ranges. Towards this end, a number of algorithms have been developed to perform this task that follow different approaches. A popular approach is based on the energy balance principle, i.e., the energy emitted in the infrared (IR; i.e., $5-1000\,\mu m$) is equal to the energy absorbed in the UV/optical wavelengths, e.g. CIGALE \citep[Code Investigating GALaxy Emission;][]{Burgarella2005, Noll2009, Boquien2019}, ProSpect \citep{Robotham2020}, MAGPHYS \citep{Cunha2008}, Prospector \citep{Leja2017}, BAGPIPES \citep{Carnall2018}.

As mentioned, AGN play an important role in galaxy evolution and their presence affects many parts of the electromagnetic spectrum. Thus, a number of the aforementioned SED fitting algorithms, have added an AGN component in the fitting process (e.g. CIGALE, ProSpect) to separate AGN and galaxy emission. SED algorithms that don't include an AGN SED component and only account for low luminosity/obscured AGN are biased against luminous/unobscured sources and underestimate the contribution of the AGN IR emission to the total IR galaxy emission. 

In a recent paper, \cite{Yang2020} presented a new branch of the CIGALE code, named X-CIGALE. Compared to CIGALE, X-CIGALE is supplemented with the modelling of AGN X-ray emission and the inclusion of polar dust. Polar dust accounts for extinction of ultraviolet (UV) and optical radiation, that is commonly found, in particular, in X-ray selected AGN \citep{Bongiorno2012}. SED fitting algorithms that include X-ray information could be instrumental in the exploitation and interpretation of the large datasets that X-ray surveys will provide.

The goal of this paper is to use the new capabilities of the X-CIGALE code on one of the largest X-ray samples available \citep[XMM-XXL;][]{Pierre2016}. XXL offers a significantly wider luminosity baseline that extents to higher luminosities compared to the fields studied in \cite{Yang2020}. Additionally, the size of the database allows to draw statistically robust conclusions in our tests. The physical properties of the XXL AGN and their host galaxies, using the CIGALE code, have been the topic of previous studies \citep[e.g.][Masoura et al. submitted]{Masoura2018, Masoura2020}. In this work, we shall focus on the effect of the new features of X-CIGALE on important SED fitting parameters. Our main purpose, is to examine how reliably the algorithm connects the X-ray flux with the UV luminosity and the rest of the other wavelengths, how accurately X-CIGALE can reproduce the X-ray properties of the AGN  and what improvements the new additions bring in its efficiency on the SED decomposition. 

 
\section{Data}
\label{sec_data}

In this section, we describe the X-ray AGN sample used in our analysis and the methodology we follow to obtain optical and IR identifications. 

\subsection{The X-ray AGN sample}
\label{sec_xdata}

Throughout our work, we use spectroscopic X-ray AGN from the XMM-XXL field \citep{Pierre2016}. XXL  is  an  international  project  based  around  an  XMM  Very Large Programme surveying two 25 deg$^2$ extragalactic fields. It has a depth of $\sim$  6 $\times$ $10^{-15}$ erg\,cm\,$^{-2}$\,s$^{-1}$ in the [0.5-2] keV band for point-like sources, with an exposure time of about 10\,ks per XMM pointing. 8445 X-ray sources have been detected in the equatorial subregions \citep[XMM-XXL-N;][]{Liu2016}. 5294 of them have SDSS counterparts. Spectroscopic redshifts are available for 2512 AGN \citep{Menzel2016} from SDSS-III/BOSS \citep{Eisenstein2011, Smee2013, Dawson2013}. These spectroscopic sources are used in our analysis. Their redshift distribution is presented in Fig. \ref{redz_distrib}.

\subsection{IR Photometry}
\label{sec_photom}

In addition to the optical (SDSS) photometry available for all our sources, we also search for counterparts in the near-IR, mid-IR and far-IR part of the spectrum. Mid-IR \citep[allWISE;][]{Wright2010} and near-IR photometry from the Visible and Infrared Survey Telescope for Astronomy \citep[VISTA]{Emerson2006} have been obtained using the Likelihood Ratio method \citep[e.g.][]{Sutherland_and_Saunders1992} as implemented in \cite{Georgakakis2011}. The process is described in detail in \cite{Georgakakis2017}. 

We also use catalogues produced by the HELP\footnote{The {\it Herschel} Extragalactic Legacy Project (HELP, http://herschel.sussex.ac.uk/) is a European funded project to analyse all the cosmological fields observed with the {\it Herschel} satellite. All the HELP data products can be accessed on HeDaM (http://hedam.lam.fr/HELP/)} collaboration to complement our mid-IR photometry with {{\it{Spitzer}} \citep{Werner2004} observations and add far-IR counterparts. HELP provides homogeneous and calibrated multiwavelength data over  the {\it Herschel}  Multitiered Extragalactic Survey \citep[HerMES,][]{Oliver2012a}  and the  H-ATLAS survey (Eales et al. 2010). The  strategy adopted by HELP is to build a master list catalog of objects  as complete as possible for each field (Shirley et al. 2019) and to use the near-IR sources from IRAC surveys as prior information for the IR maps. The  XID+ tool \citep{Hurley2017}, developed for this purpose, uses a  Bayesian probabilistic framework  and works with prior positions. At the end, a flux is  measured, in a probabilistic sense, for all the near-IR sources of the master list. The XMM-LSS field was covered by two {\it Spitzer} surveys, SpUDS \citep[Spitzer UKIDSS Ultra Deep Survey,][]{Caputi2011} and SWIRE/SERVS \citep{Lonsdale2003}. The prior positions are defined with the SpUDS and SWIRE/SERVS surveys and the fluxes are measured for the Spitzer MIPS/24 microns, and Herschel PACS and SPIRE bands. In this work, only the MIPS and SPIRE fluxes are considered, given the much lower sensitivity of the PACS observations for this field \citep{Oliver2012}.

The W1, W2 photometric bands of WISE nearly overlap with IRAC1, IRAC2 from {\it Spitzer}. When a source has been detected by both IR surveys, we only consider the photometry with the highest signal-to-noise ratio (SNR). Similarly, when both W4 and MIPS photometry is available, we consider only the latter due to the higher sensitivity of {\it Spitzer} compared to WISE. The number of sources available in each photometric band  of our final sample is presented in Table \ref{table:photometry}. Table \ref{table:sensitivity}, shows the sensitivity of each photometric band, used in our analysis.

\begin{table*}
\caption{Number of available spectroscopic sources in different photometric bands. 2342/2509 ($\sim 93\%$) of the sources, have optical and mid-IR photometry available (either W1,W2 or IRAC1, IRAC2). 1276 ($\sim 51\%$) have far-IR photometry available (and mid-IR).}
\centering
\setlength{\tabcolsep}{1.5mm}
\begin{tabular}{cccccccccc}
 \hline
{Total} & {W1, W2 / IRAC1, IRAC2} & {IRAC3} &{IRAC4}  & {W3}  & {MIPS1} & {{\it Herschel}} & {all bands} \\
       \hline
2509  & 2342 & 1122 & 1135  & 1240   & 1298 & 1276 & 408 \\
       \hline
\label{table:photometry}
\end{tabular}
\end{table*}

\begin{table*}
\caption{5$\sigma$ sensitivity (mag) of the photometric bands, used in our analysis.}
\centering
\setlength{\tabcolsep}{1.mm}
\begin{tabular}{cccccc}
 \hline
{u, g, r, i, z} & {J, H, K} & {W1, W2, W3, W4} & {IRAC1, IRAC2, IRAC3, IRAC4} & {MIPS1} & {SPIRE} \\
       \hline
22.15, 23.13, 22.70, 22.20, 20.71  & 20.6, 19.8, 18.5 & 16.83, 15.60, 11.32, 8.0 & 24.09, 23.66, 21.90, 21.85  & 19.8   & 13.9, 13.5, 13.2 \\
       \hline
\label{table:sensitivity}
\end{tabular}
\end{table*}

\section{Analysis}
\label{sec_analysis}
X-CIGALE requires the intrinsic, i.e. unabsorbed, flux of the X-ray AGN. In this section, we describe how we estimate the X-ray absorption of the sources to infer their intrinsic X-ray flux. We also describe the modules and parameters used for the SED fitting process and present our final sample. 

\subsection{Estimation of the X-ray properties}
\label{sec_xproperties}

To estimate the intrinsic, X-ray flux of each source, we need to measure their hydrogen column density, $\rm N_H$, that quantifies the X-ray absorption. Towards this end, we use the number of photons in the soft ($0.5-2.0$\,keV) and the hard ($2.0-8.0$\,keV) bands that are provided in the \cite{Liu2016} catalogue. We then apply a Bayesian approach to estimate the hardness ratio, $\rm HR=\frac{H-S}{H+S}$, of each source, where H and S are the counts in the soft and hard bands, respectively. Specifically, we utilise the Bayesian Estimation of Hardness Ratios code \citep[BEHR;][]{Park2006}, that accounts for the Poissonian nature of the observations. These HR values are then inserted in the PIMMS tool \citep[Portable, Interactive, Multi-Mission Simulator;][]{Mukai1993} to estimate the $\rm N_H$ of each source. In this process, we assume that the power law of the X-ray spectra have a fixed photon index, $\Gamma =1.8$ and the value of the galactic $\rm N_H$ is $\rm N_H =10^{20.25}\,cm^{-2}$. The distribution of $\rm N_H$ of our AGN sample is presented in Fig. \ref{nh_distrib}. Due to the Bayesian nature of our calculations, some $\rm log\, N_H$ values are below 20.25, i.e. the galactic absorption. 

The X-ray absorption that was estimated through PIMMS and based on which the intrinsic X-ray flux is inferred, does not necessarily correlate with the dust that the observed mid- and far-IR emission exhibits. Although there are studies that have found a correlation between optical/IR obscuration and X-ray absorption \citep[e.g.][]{Civano2012}, these correlation presents a large scatter \citep[e.g.][]{Jaffarian2020}. This scatter is attributed to e.g. i) the X-ray column density variability \citep[e.g.][]{Reichert1985, Yang2016}, ii) absorbing material located at galactic scales \citep[e.g.][]{Malizia2020} that is mainly heated by star formation rather than AGN, iii) the fact that different obscuration criteria are sensitive to different amounts of obscuration \citep[e.g.][]{Masoura2020} and iv) a dust-to-gas ratio that is different from the Galactic. Therefore, a source may present X-ray absorption without being optical red (dust-free gas), may also be heavily X-ray absorbed with broad UV/optical lines \citep[e.g.][]{Li2019}, can be optically red, with absorbed AGN emission in its SED without being X-ray absorbed \citep[e.g.][]{Masoura2020} and can present higher X-ray absorption than what is expected from its optical extinction. Thus, we do not necessarily expect consistency between the X-ray absorption estimated through PIMMS with that from dust, estimated by X-CIGALE, by modelling the mid and far-IR emission.

\begin{figure}
\centering
  \includegraphics[width=1.\linewidth]{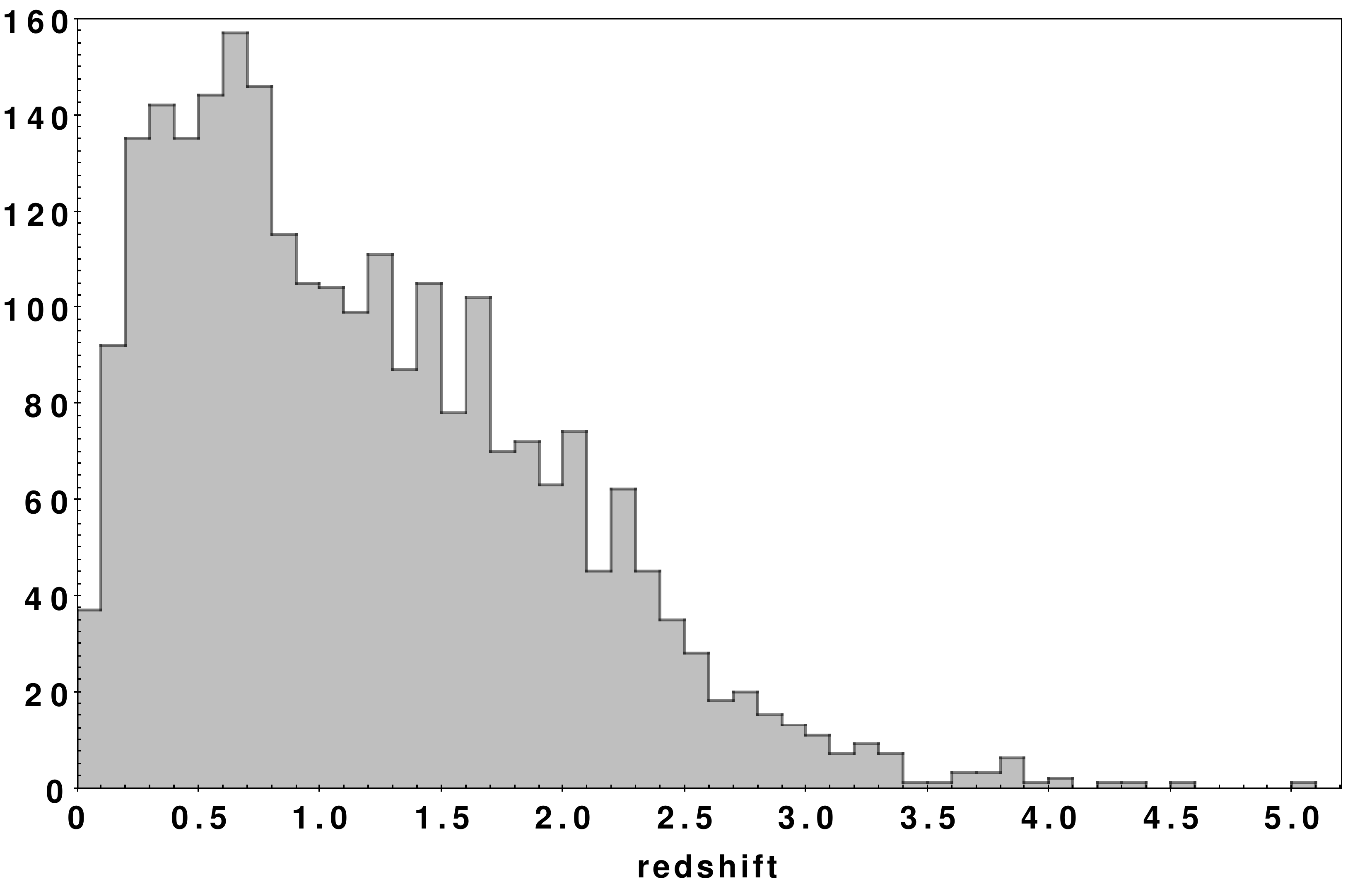}
  \caption{The redshift distribution of the X-ray AGN sample.}
  \label{redz_distrib}
\end{figure}

\begin{figure}
\centering
  \includegraphics[width=1.\linewidth]{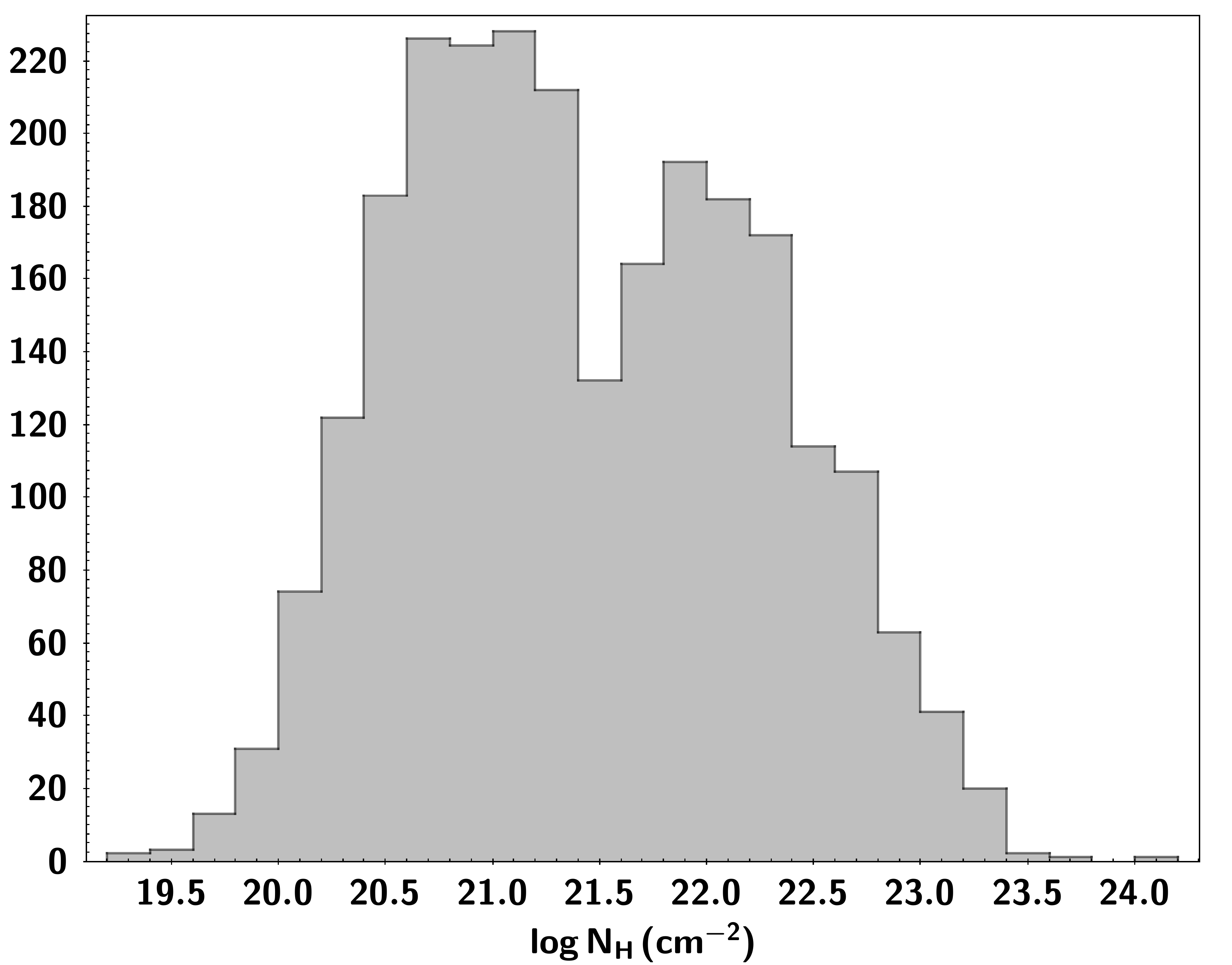}
  \caption{Distribution of the X-ray absorption, $\rm N_H$, of the X-ray AGN sample.}
  \label{nh_distrib}
\end{figure}

\begin{table*}
\caption{The models and the values for their free parameters used by X-CIGALE for the SED fitting of our galaxy sample. For the definition of the various parameter see section \ref{sec_sed_analysis}.} 
\centering
\setlength{\tabcolsep}{1.mm}
\begin{tabular}{cc}
       \hline
Parameter &  Model/values \\
	\hline
\multicolumn{2}{c}{Star formation history: delayed model} \\
e-folding time & 100, 500, 1000, 5000 \\ 
Stellar age age & 500, 1000, 3000, 5000, 7000 \\
\hline
\multicolumn{2}{c}{Simple Stellar population: Bruzual \& Charlot (2003)} \\
Initial Mass Function & Chabrier (2003)\\
Metallicity & 0.02 (Solar) \\
\hline
\multicolumn{2}{c}{Galactic dust extinction} \\
Dust attenuation law & Calzetti et al. (2000) \\
Reddening $E(B-V)$ & 0.0, 0.1, 0.2, 0.3, 0.4, 0.5, 0.6, 0.7, 0.8, 0.9 \\ 
\hline
\multicolumn{2}{c}{Galactic dust emission: Dale et al. (2014)} \\
$\alpha$ slope in $dM_{dust}\propto U^{-\alpha}dU$ & 1.0, 1.5, 2.0, 2.5, 3.0 \\
\hline
\multicolumn{2}{c}{AGN module: SKIRTOR)} \\
Torus optical depth at 9.7 microns $\tau _{9.7}$ & 7.0 \\
Torus density radial parameter p ($\rho \propto r^{-p}e^{-q|cos(\theta)|}$) & 1.0 \\
Torus density angular parameter q ($\rho \propto r^{-p}e^{-q|cos(\theta)|}$) & 1.0 \\
Angle between the equatorial plan and edge of the torus & $40^{\circ}$ \\
Ratio of the maximum to minimum radii of the torus & 20 \\
Viewing angle  & $30^{\circ}\,\,\rm{(type\,\,1)},70^{\circ}\,\,\rm{(type\,\,2)}$ \\
AGN fraction & 0.0, 0.01, 0.1, 0.2, 0.3, 0.4, 0.5, 0.6, 0.7, 0.8, 0.9, 0.99 \\
Extinction law of polar dust & SMC \\
$E(B-V)$ of polar dust & 0.0, 0.01, 0.02, 0.03, 0.05, 0.1, 0.2, 0.4, 0.6, 1.0, 1.8 \\
Temperature of polar dust (K) & 100 \\
Emissivity of polar dust & 1.6 \\
\hline
\multicolumn{2}{c}{X-ray module} \\
AGN photon index $\Gamma$ & 1.8 \\
Maximum deviation from the $\alpha _{ox}-L_{2500 \AA}$ relation & 0.2 \\
LMXB photon index & 1.56 \\
HMXB photon index & 2.0 \\
\hline
\label{table_X-CIGALE}
\end{tabular}
\end{table*}

\subsection{SED fitting with X-CIGALE}
\label{sec_sed_analysis}

The fitting capabilities of CIGALE  have been recently extended to X-rays with the development of X-CIGALE \citep{Yang2020}, in order to improve the characterisation of the AGN component. The X-ray emission is connected to the AGN emission at other wavelengths via the $\alpha_{\rm ox}-L_{2500\AA}$ relation of \cite{Just2007}, where $L_{2500\AA}$ is the intrinsic (de-reddened) UV  luminosity and $\alpha_{\rm ox}$ the spectral slope between UV($2500\AA$) and X-ray (2 keV), $\alpha _{ox}=-0.3838\,\rm {log}\,(L_{2500\AA}/L_{2keV})$. The contribution of X-ray binaries is also considered and modelled as a function of SFR and stellar mass of the host galaxy. The clumpy two-phase torus model, SKIRTOR, based on 3D radiation-transfer \citep{Stalevski2012, Stalevski2016} is used for the  UV to far-IR emission of the AGN with some modifications keeping the energy balance: the original emission of the accretion disc is updated with the spectral energy distribution of \cite{Feltre2012} and dust extinction and emission in the poles of type 1 AGN is also considered \citep[e.g.][]{Bongiorno2012, Tristram2014, Asmus2014, Asmus2019}. We refer to \cite{Yang2020} for a full description of X-CIGALE. Here, we describe the main steps to build the models and fit our X-ray to far-IR data. The modules and input parameters we use in our analysis are presented in Table \ref{table_X-CIGALE}.

\subsubsection{Galaxy emission}
For the sake of simplicity and since we will not study in detail the SFR properties of our sources, we adopt a simple star formation history (SFH).  The galaxy component is built using a delayed SFH ($\rm SFR\propto t \times \exp(-t/\tau)$).  We checked that the addition of a recent burst \citep[e.g.][]{Masoura2018, Malek2018} did not change our results. The stellar emission is modelled using the \cite{Bruzual_Charlot2003} template. A \cite{Chabrier2003} initial mass function (IMF) is used with metallicity equal to 0.02. The stellar emission is attenuated with the \cite{Calzetti2000} attenuation law. The IR SED of the dust heated by stars is implemented with the \cite{Dale2014} template. 

\subsubsection{AGN emission}
\label{sec_agn_emission}
The AGN emission is modelled using the SKIRTOR template \citep{Stalevski2012, Stalevski2016}. A polar dust component is added that is modelled as a dust screen absorption and a grey-body emission. As in \cite{Yang2020}, we adopt the SMC extinction curve \citep[Small Magellanic Cloud;][]{Prevot1984} with  $\rm E_{(B-V)}$ as a free input parameter (see section 5 for the effects of $\rm E_{(B-V)}$ on SED fitting). The grey-body dust re-emission  is parametrised  with a temperature of 100 K and an emissivity index of 1.6.  This emission is supposed to be isotropic and thus contributes to the IR emission of both type 1 and type 2 AGNs. We will discuss in section \ref{sec_extinction_curve} the effect of a modification of the dust temperature and of the  extinction curve. The contribution of the AGN to the total SED is quantified by the AGN fraction, frac$\rm _{AGN}$, defined as the fraction of the total IR emission coming from the AGN. Following \cite{Yang2020} two viewing angles are considered, 30$\rm^o$ and 70$\rm^o$, for type 1 and type 2 AGN, respectively. 

The photon index  $\Gamma$ of the AGN X-ray spectrum is fixed to 1.8, for consistency with the value used for the estimation of $\rm N_H$ (see section \ref{sec_xproperties}). We adopt a  maximal acceptable value $\lvert \Delta \alpha_{\rm ox} \rvert_{\rm max}$=0.2 for the  dispersion  of the $\alpha_{\rm ox}-L_{2500\AA}$ relation \citep{Risaliti2017}. This value is also adopted in \cite{Yang2020} and corresponds to $\approx 2\sigma$ scatter in the $\alpha _{ox}-\rm L_{2500\,\AA}$ relation \citep{Just2007}. Nine values of $\alpha_{\rm ox}$ are defined (from -1.9 to -1.1 with a step of 0.1) and  X-CIGALE adds an X-ray flux  to each (UV to far-IR AGN SED,  $\alpha_{\rm ox}$) pair, multiplying the number of AGN models by 9.

\begin{table*}
\caption{Number of AGN that satisfy our selection criteria (section \ref{sec_finalsample}) for different configurations of the SED fitting process.}
\centering
\setlength{\tabcolsep}{1.5mm}
\begin{tabular}{ccccc}
 \hline
{Total} & {$\rm f_X$, polar dust} & {no $\rm f_X$, polar dust}  & {$\rm f_X$, no polar dust}  & {no $\rm f_X$, no polar dust}\\
       \hline
2509  & 2348 & 2274 & 2337  & 2274 \\
       \hline
\label{table:samples}
\end{tabular}
\end{table*}

\begin{figure*}
\centering
\begin{subfigure}{.5\textwidth}
  \centering
  \includegraphics[width=1.\linewidth]{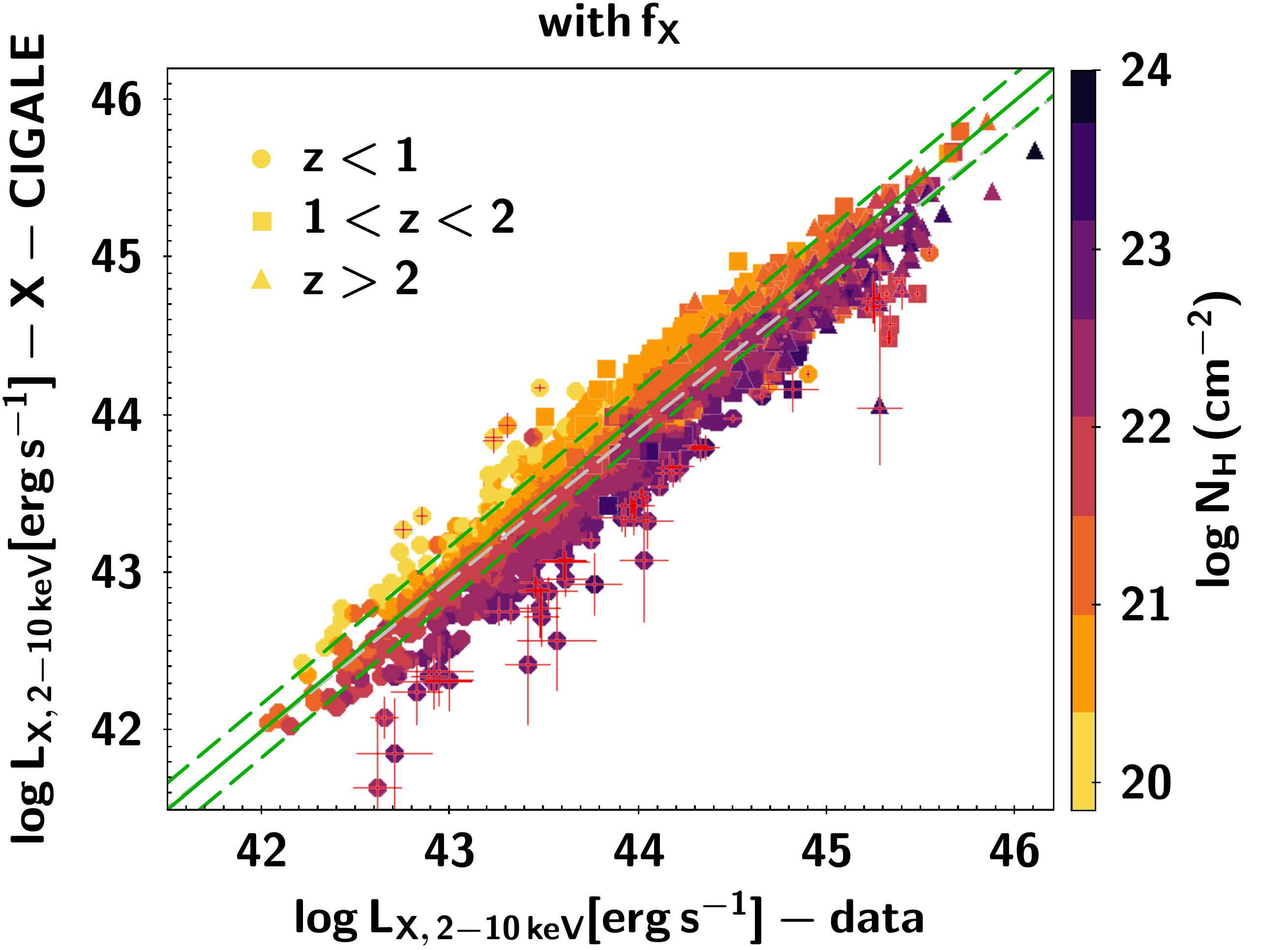}
  \label{lx_comp_fx}
\end{subfigure}%
\begin{subfigure}{.5\textwidth}
  \centering
  \includegraphics[width=1.\linewidth]{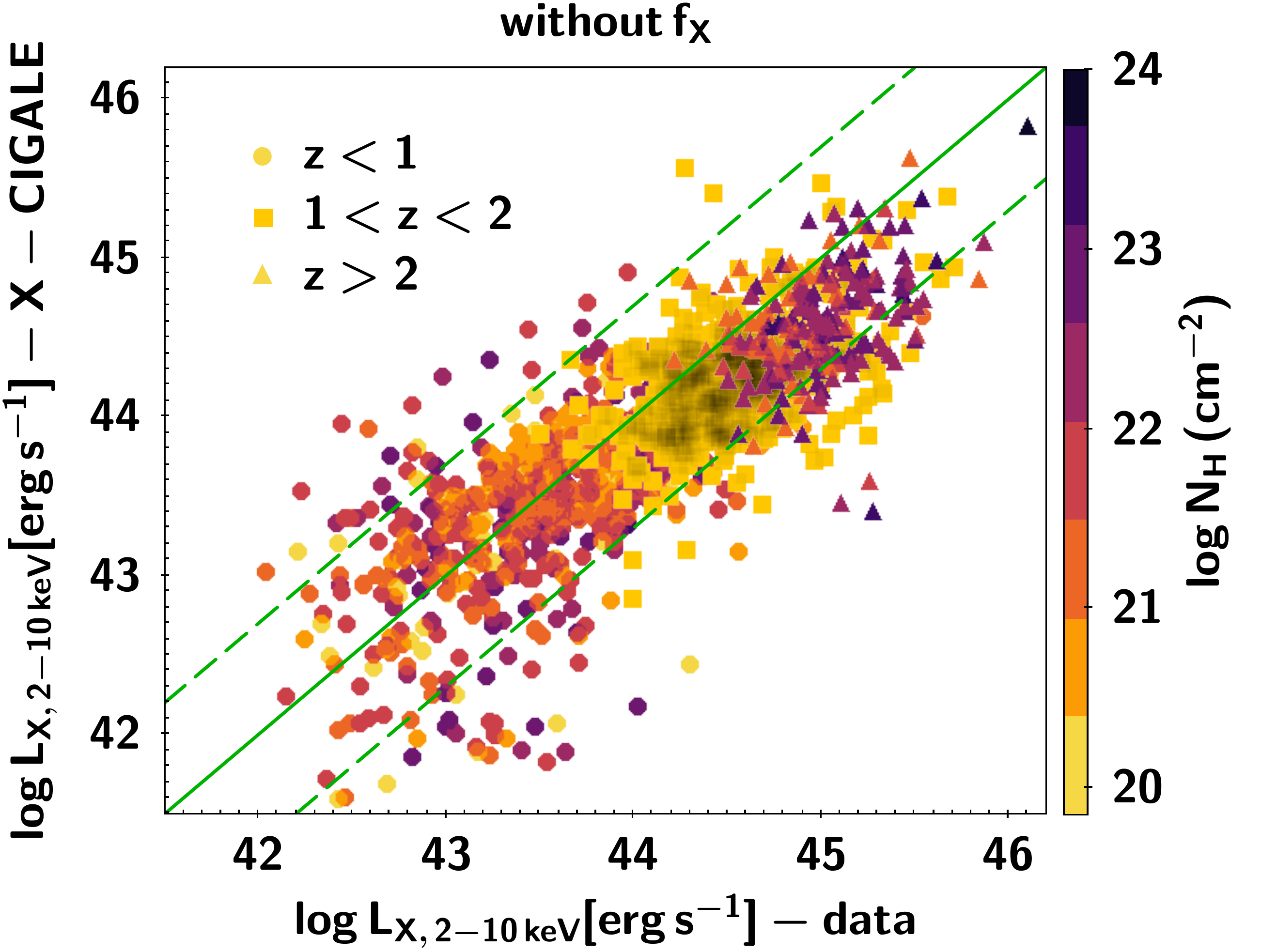}
  \label{lx_comp_nofx}
\end{subfigure}
\caption{Comparison of the intrinsic, X-ray luminosity in the 2-10\,keV band estimated by the SED fitting with that from the input catalogue. Green, solid and dashed lines present the 1:1 correspondence and the errors added in quadrature of the X-ray luminosity calculations from X-CIGALE and those quoted in the X-ray catalogue \citep{Menzel2016}, respectively. The grey, dashed line shows the $\chi ^2$ fit of the calculations. Symbols are colour coded, based on their N$\rm H$ values. Left panel: Results with the X-ray flux included in the SED. X-CIGALE calculations are consistent with those from the input catalogue, at all luminosities spanned by our sample. There are 70 sources that their L$_{\rm X}$ calculation differs by more than 0.5\,dex from their input values (shown with errorbars, see text for more details). Right panel: The X-ray flux is not included in the SED. In this case, the scatter is significantly larger while the error of the $\rm L_X$ calculations from X-CIGALE, increases by $\approx 4\times$.}
\label{lx_comp}
\end{figure*}

\begin{figure}
\centering
  \includegraphics[width=1.\linewidth]{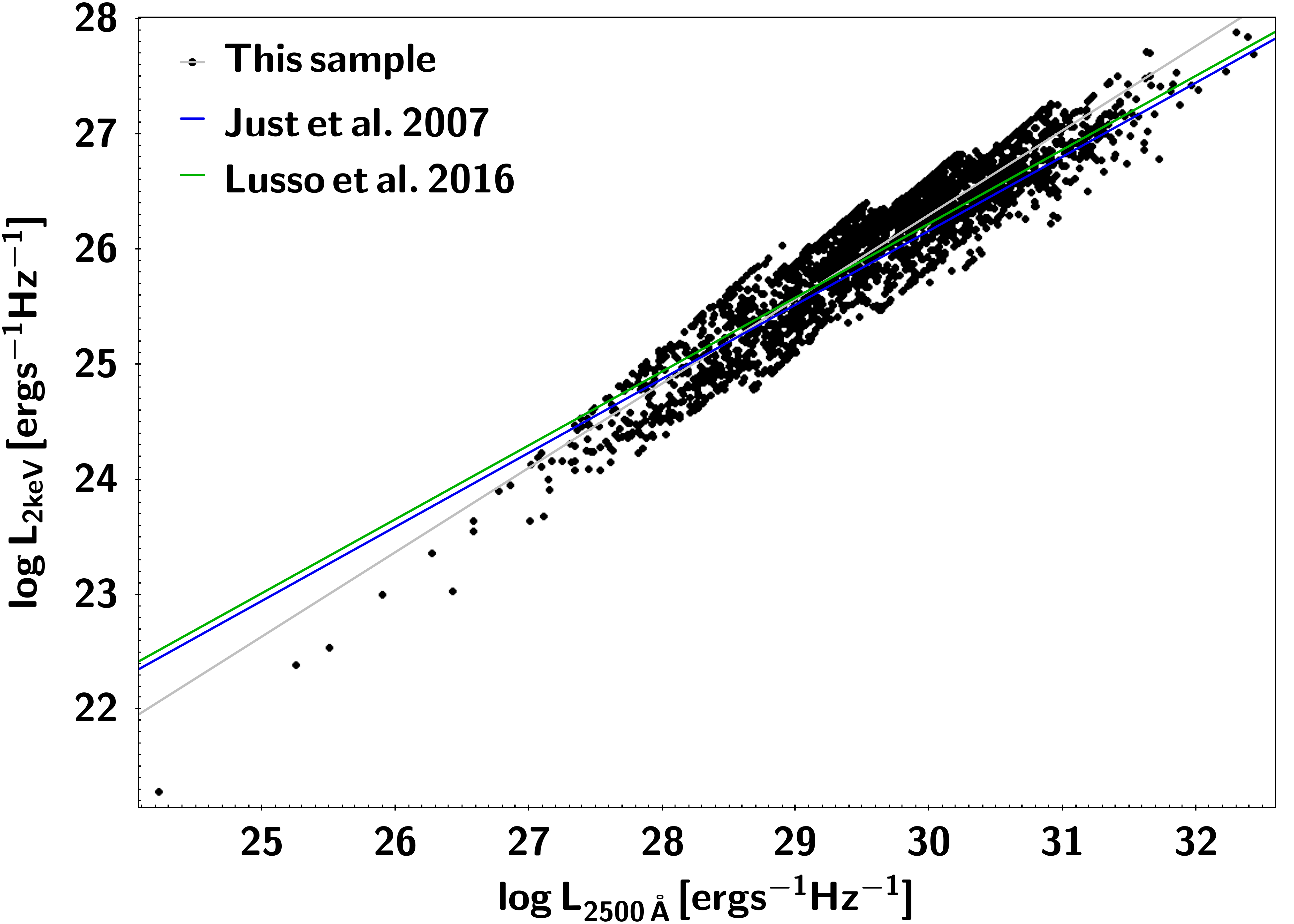}
  \label{ebv_data_mock_nofxebv}
\caption{$\rm L_{2\,keV}$ vs. $\rm L_{2500\,\AA}$ relation. Blue line presents the \cite{Just2007} relation, that X-CIGALE uses to connect the X-ray flux with the UV (2500 \AA) luminosity. Green line shows the $\rm L_{2\,keV}$ vs. $\rm L_{2500\,\AA}$ from \cite{Lusso2016}. For comparison, we also plot the fit from our calculations (grey line).}
\label{just_relation}
\end{figure}

\subsubsection{SED fitting and parameter estimation}
The spectral models described above are normalised to the creation of 1 $\rm M_{\sun}$. The first step of the fitting process is to scale each model to the observations by minimising its $\chi^2$  \citep{Noll2009, Boquien2019}. After this scaling operation  extensive quantities as luminosities are defined, including  the intrinsic $L_{2500\AA}$ which is an output of the SKIRTOR module:  models which do not satisfy the  $\alpha_{\rm ox}-L_{2500\AA}$ relation and its maximum dispersion  are discarded. The likelihood of the remaining models are computed  and parameter values are estimated from their marginalised probability distribution function (likelihood weighted mean and standard deviation). The best model corresponding to each observed SED is also an output of the code.  

\subsubsection{Mock catalogues}
\label{sec_mock}
The validity of a parameter estimation can be assessed through the analysis of a mock catalogue. When this option is chosen, the code considers the best fit of each object and a mock catalogue is built. Each best flux being modified  by injecting noise taken from a Gaussian distribution with the same standard deviation as the observed flux. The mock data are then analysed in the same way as the observed data and the accuracy of the parameter estimation can be tested by comparing input (ground truth) and output (estimated) values. Our tests using the results from the mock catalogues are presented in the Appendix.

\subsection{Final sample}
\label{sec_finalsample}

In our analysis, we compare the estimations of the $\rm frac_{AGN}$, namely the value of the best model and then mean of the probability density function (PDF) distribution, to select the X-ray AGN with the most reliable SED measurements. For that purpose, we apply the following criteria: we exclude sources for which their best AGN fraction values are zero while their Bayesian AGN fraction is greater than 0.4. Such a large difference between the best and the Bayesian values, is a strong indicator that the code failed to accurately fit the SED since the PDF distribution is not consistent with the best model. We also exclude AGN with reduced $\chi ^2$, $\chi ^{2}_{\rm red}>10$ from the SED fitting process. These two criteria exclude $\sim 4-7\%$ of our sources. Finally, we remove from our final sample, sources with $\rm {L_{X}<10^{42}\, erg\,s^{-1}}$ to minimise contamination from inactive galaxies. Table \ref{table:samples} presents the number of sources that satisfy our selection criteria for each configuration of the SED fitting process used in our analysis.

\begin{figure}
\centering
\begin{subfigure}{.27\textwidth}
  \centering
  \includegraphics[width=1.\linewidth]{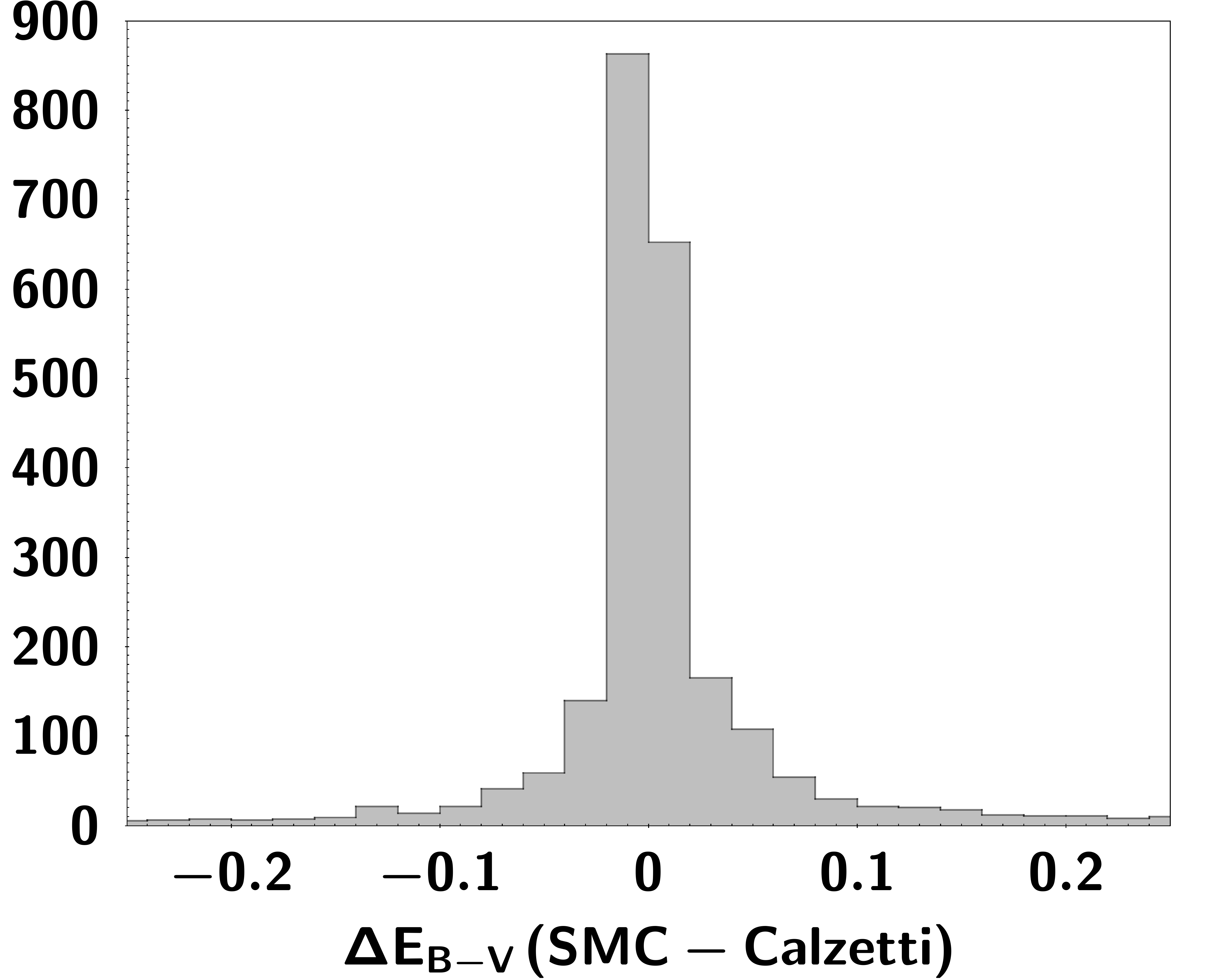}
  \caption{}
  \label{fig_ebv_smc_vs_calzetti}
\end{subfigure}
\begin{subfigure}{.27\textwidth}
  \centering
  \includegraphics[width=1.\linewidth]{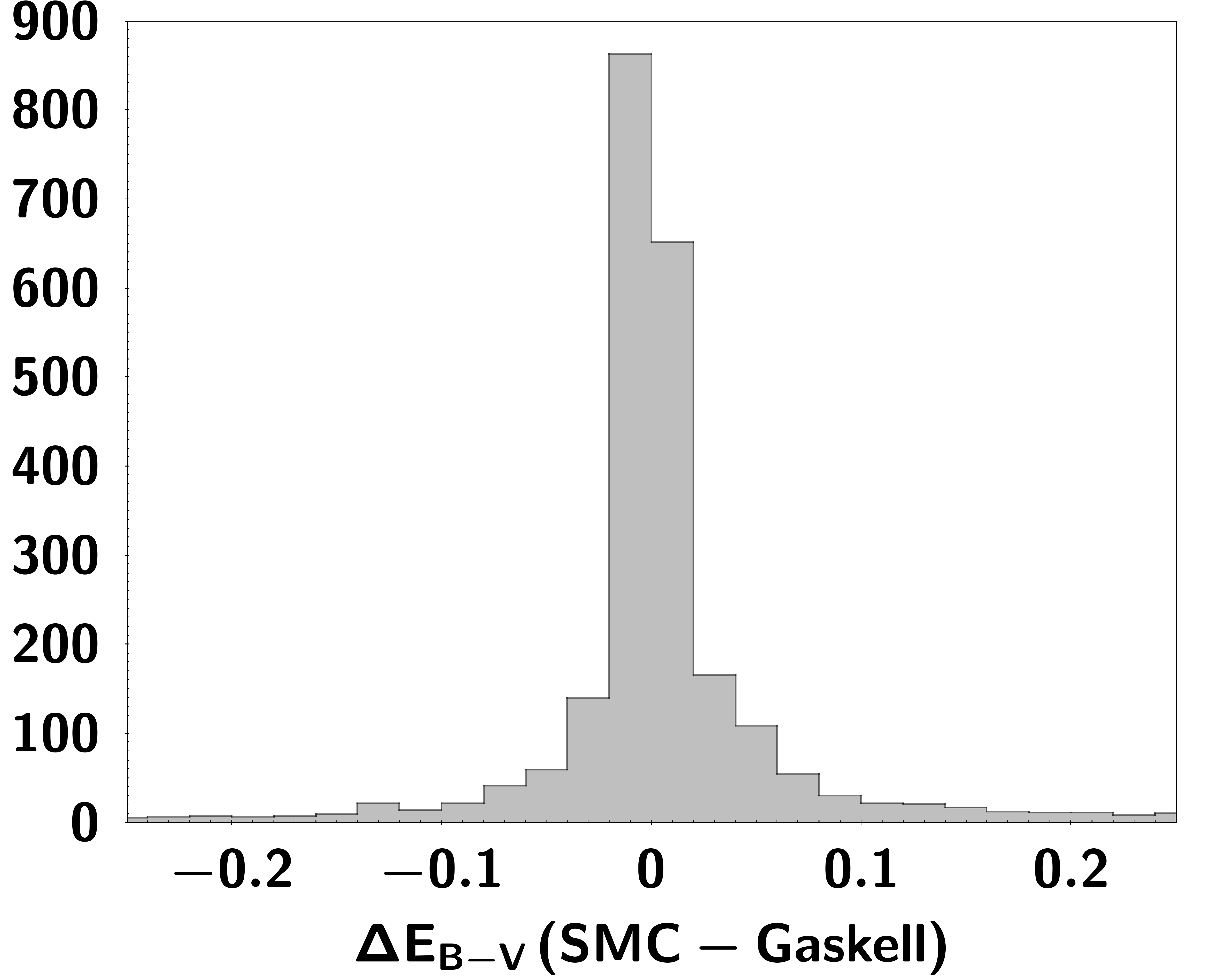}
  \caption{}
  \label{fig_ebv_smc_vs_gaskell}
\end{subfigure}
\begin{subfigure}{.27\textwidth}
  \centering
  \includegraphics[width=1.\linewidth]{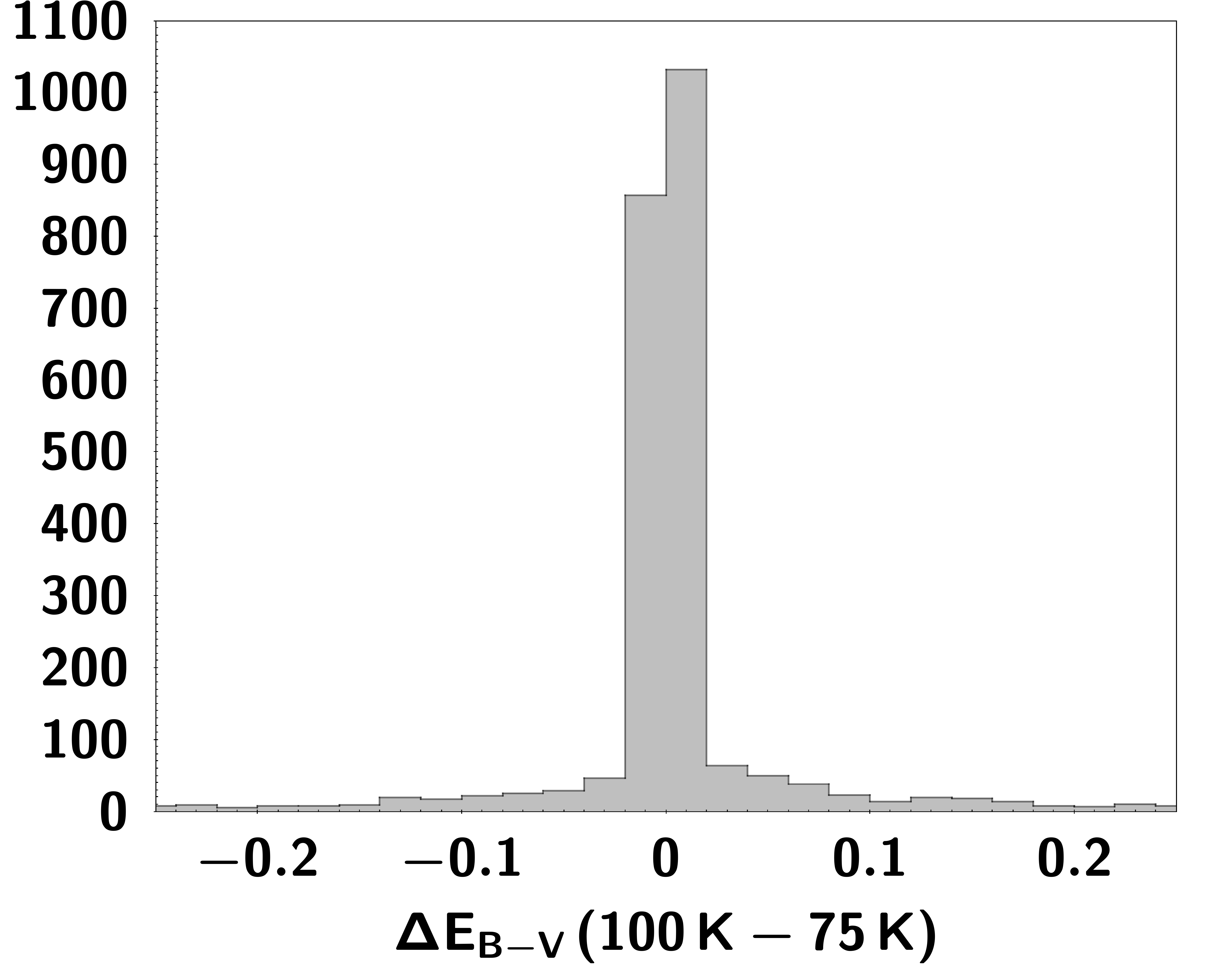}
  \caption{}
  \label{fig_ebv_100_vs_75}
\end{subfigure}
\begin{subfigure}{.27\textwidth}
  \centering
  \includegraphics[width=1.\linewidth]{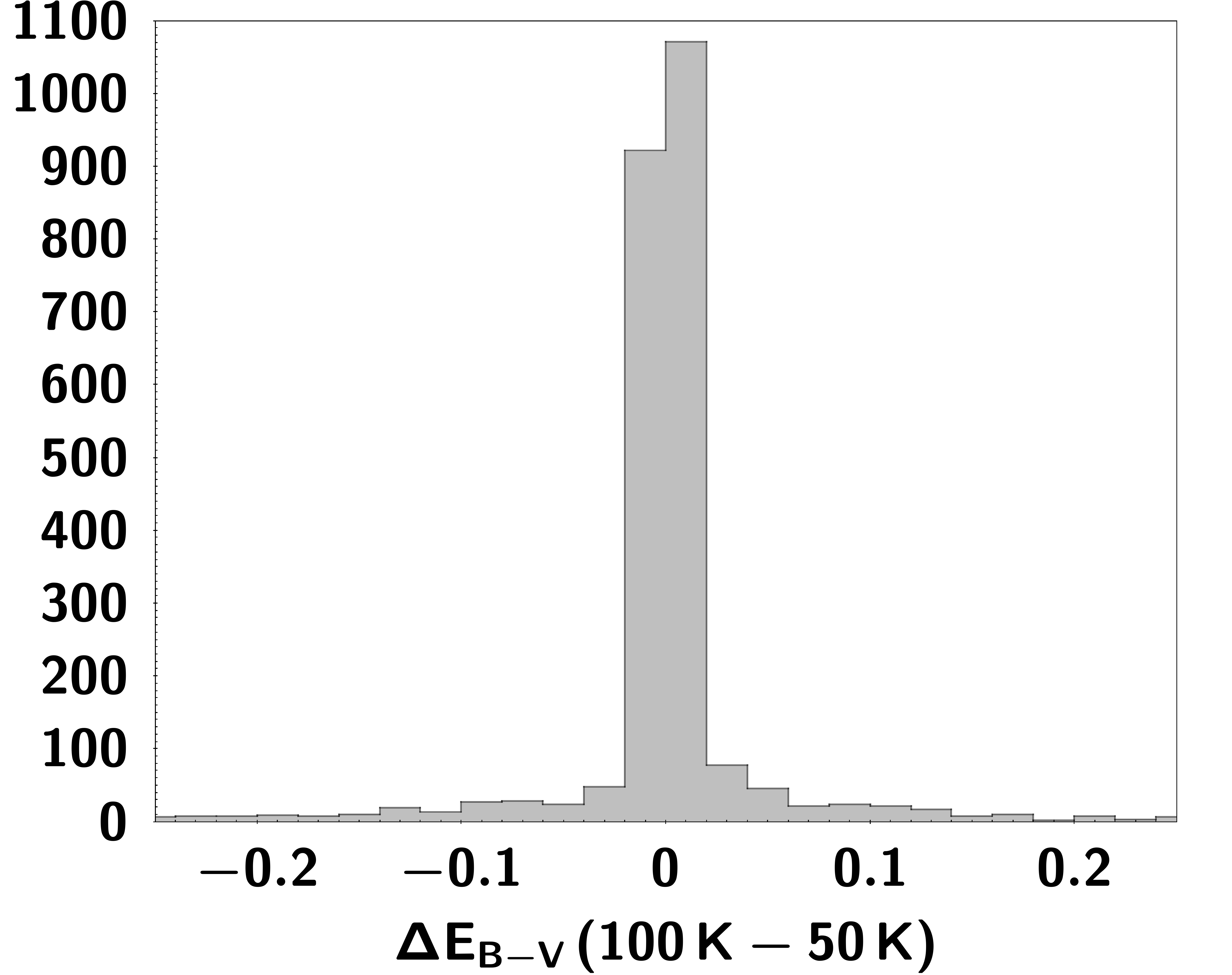}
  \caption{}
  \label{fig_ebv_100_vs_50}
\end{subfigure}
\caption{Distribution of the difference of polar dust measurements, for different extinction laws (panels a, b) and grey-body dust temperatures (panels c, d). All distributions are highly peaked at zero, which indicates that polar dust calculations are not sensitive to the choice of these parameters.} 
\label{frac_plot_mock_data_lx}
\end{figure}

\begin{figure}
\centering
\begin{subfigure}{.27\textwidth}
  \centering
  \includegraphics[width=1.\linewidth]{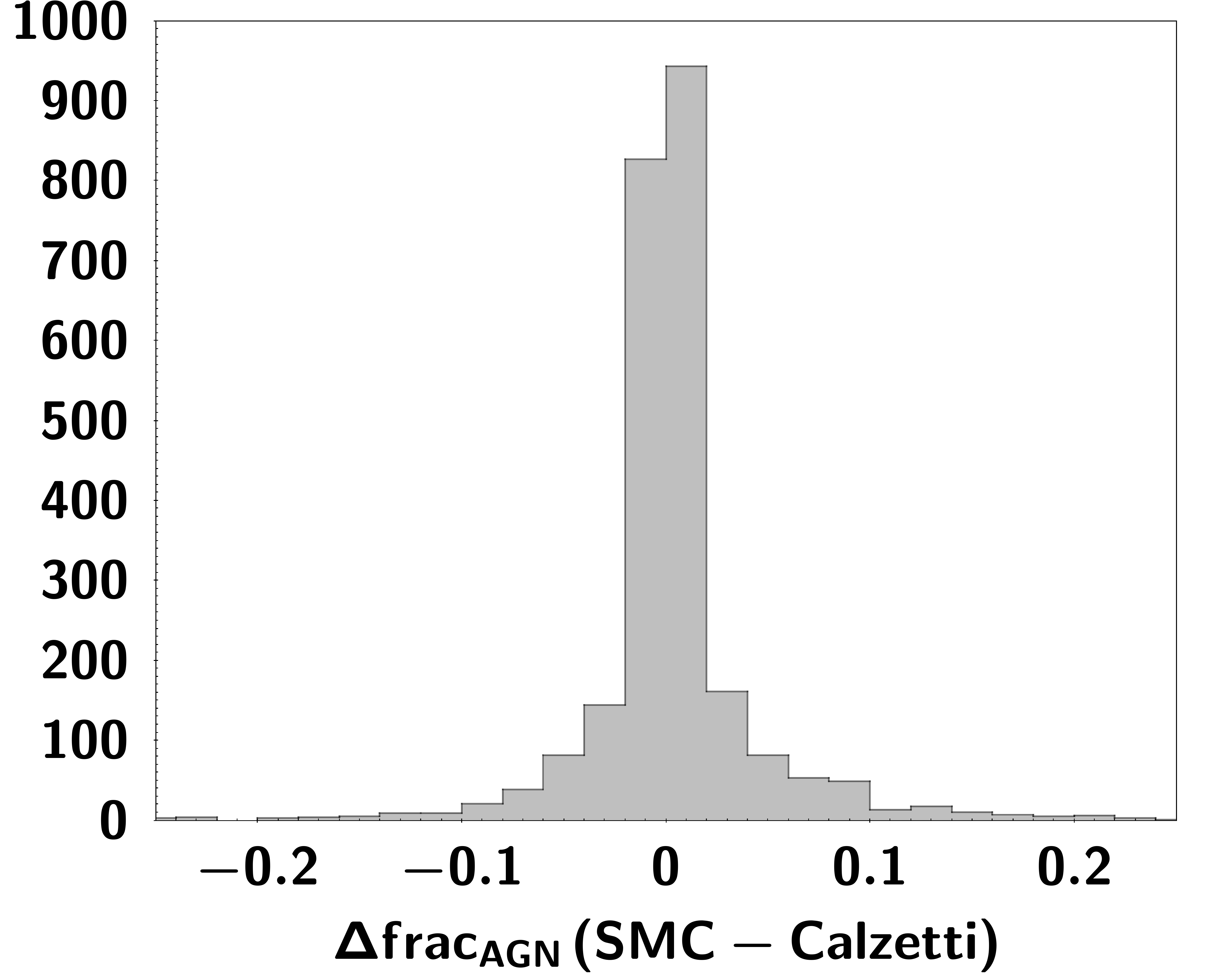}
  \caption{}
  \label{fig_frac_smc_vs_calzetti}
\end{subfigure}
\begin{subfigure}{.27\textwidth}
  \centering
  \includegraphics[width=1.\linewidth]{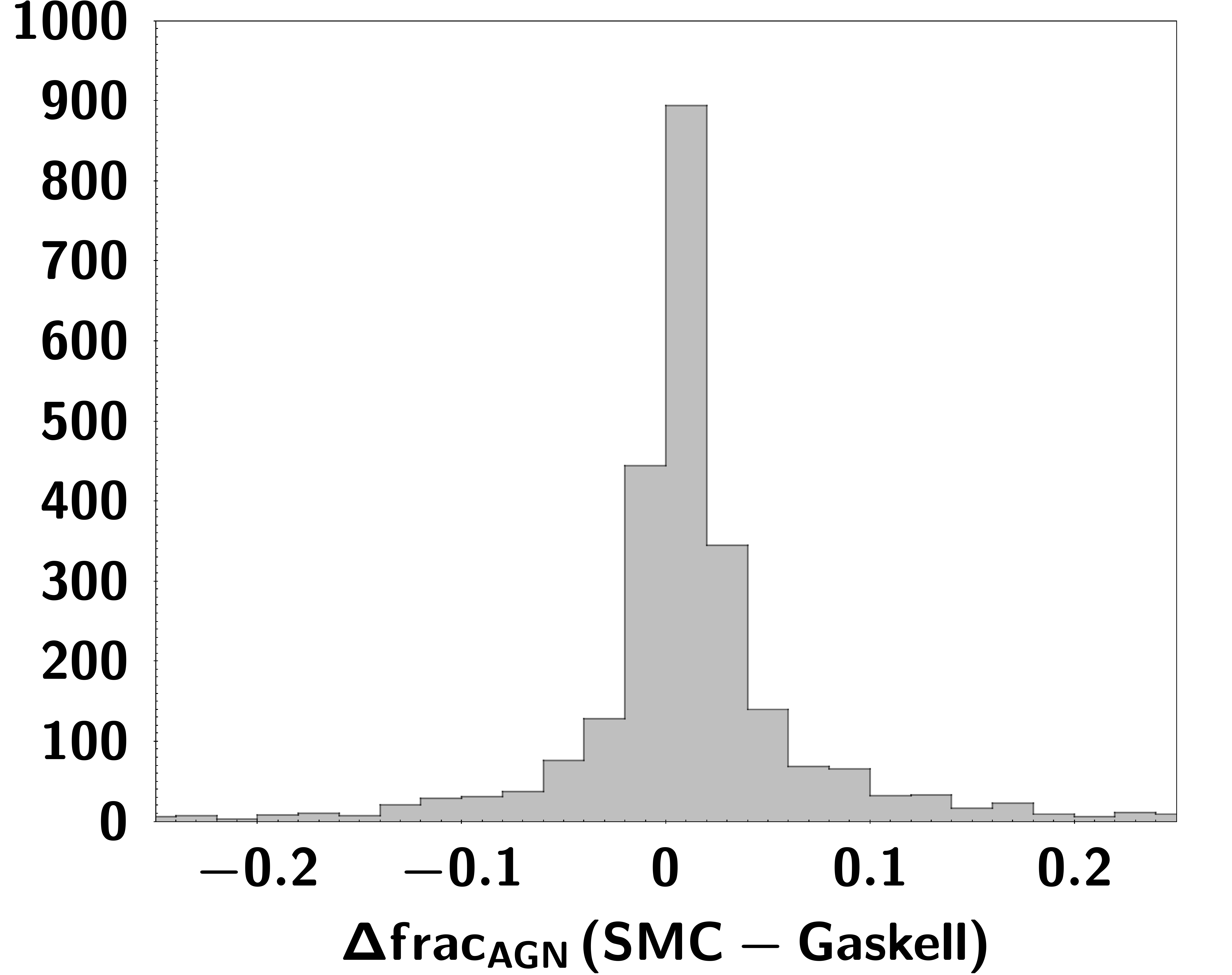}
  \caption{}
  \label{fig_frac_smc_vs_gaskell}
\end{subfigure}
\begin{subfigure}{.27\textwidth}
  \centering
  \includegraphics[width=1.\linewidth]{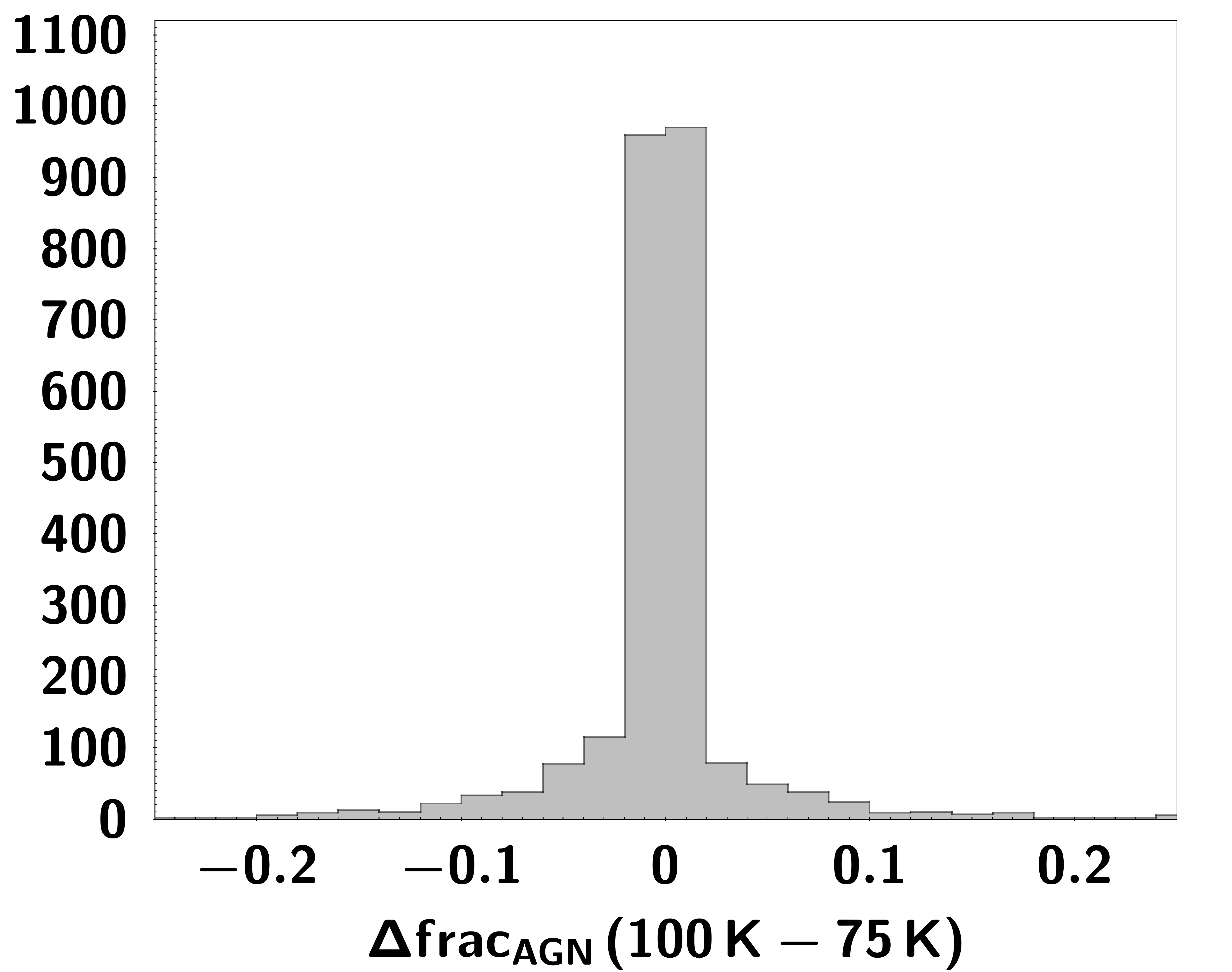}
  \caption{}
  \label{fig_frac_100_vs_75}
\end{subfigure}
\begin{subfigure}{.27\textwidth}
  \centering
  \includegraphics[width=1.\linewidth]{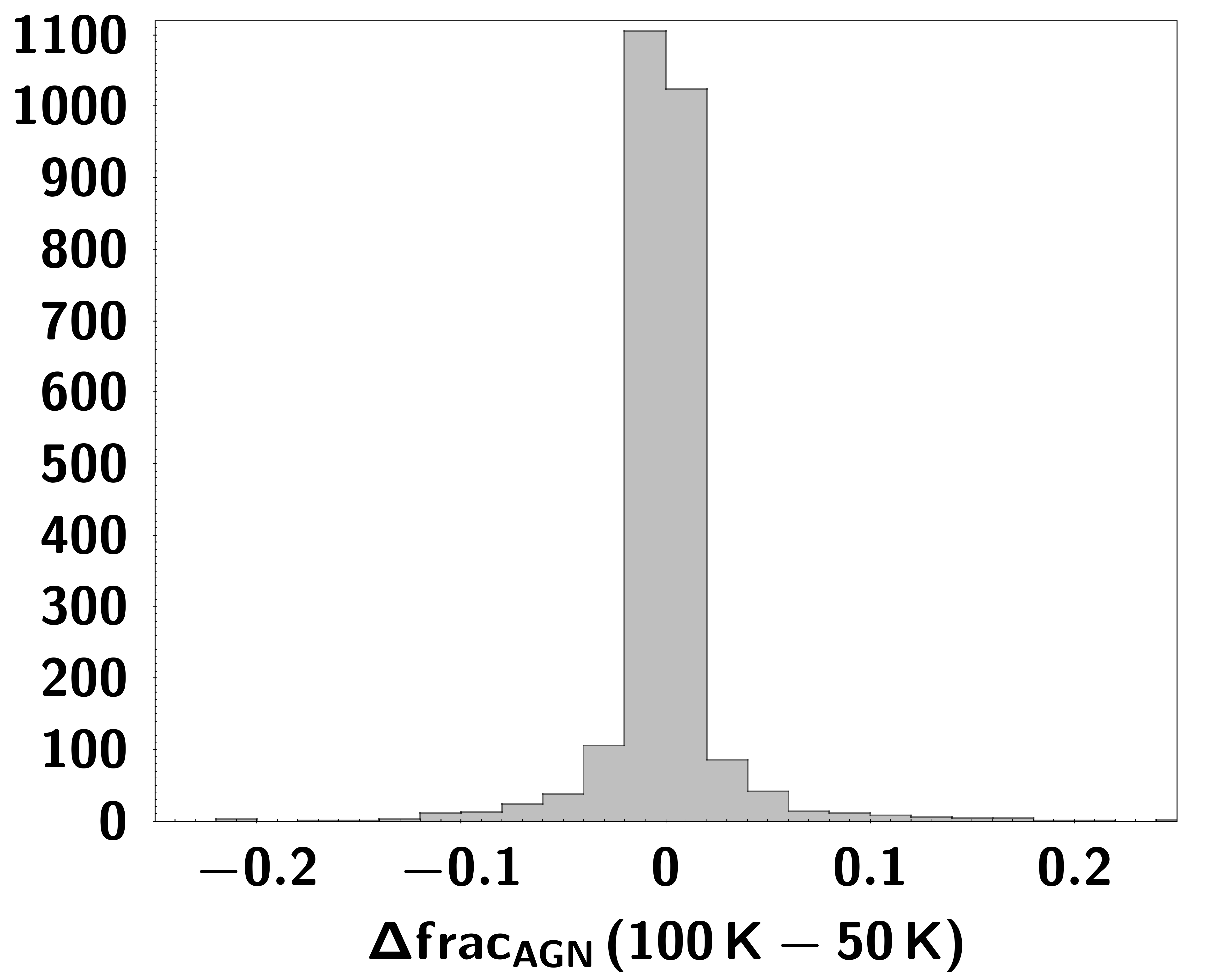}
  \caption{}
  \label{fig_frac_100_vs_50}
\end{subfigure}
\caption{Same as in the previous plot, but for the AGN fraction. Different extinction laws (panels a, b) and grey-body dust temperatures (panels c, d) do not affect the $\rm frac _{AGN}$ measurements.}
\label{frac_plot_mock_data_lx}
\end{figure}

\section{X-CIGALE performance}
\label{xcigale_reliability}

\subsection{X-ray luminosity}
\label{sec_lx}

First, we compare the intrinsic, X-ray luminosities of the AGN estimated by X-CIGALE with those from the input catalogue. The results are presented in Fig. \ref{lx_comp}. X-CIGALE estimates are consistent with those from the input catalogue, at all luminosities spanned by our sample. A $\chi ^2$ fit (grey, dashed line) gives $\rm log\,L_{X, X-CIGALE}=(0.962\pm0.023)\,log\,L_{X, data}+1.694\pm0.085$. Green, dashed lines present the errors added in quadrature of the X-ray luminosity calculations from X-CIGALE and those quoted in the X-ray catalogue \citep{Menzel2016}. There are 70 outliers, i.e., sources for which their $\rm L_X$ calculations differ by more than 0.5\,dex from their input values (shown with errorbars). The vast majority of sources that X-CIGALE underestimates their L$_X$ value by more than 0.5\,dex are AGN with high X-ray absorption ($\rm N_H>10^{22.5-23.0}\,cm^{-2}$). 82\% of these sources also lack far-IR photometry and 57\% do not have IR coverage above 8$\,\rm \mu m$. Thus, a possible explanation for the discrepant $\rm L_X$ values could be that our $\rm N_H$ estimations are not accurate and/or the lack of available IR photometry does not allow X-CIGALE to properly fit these SEDs. Lowering the $\rm L_X$ difference threshold to 0.3\,dex to characterise a source as outlier, leads to the same conclusions. On the other hand, those sources that X-CIGALE overestimates their value by more than 0.5\,dex ($<0.4\%$ of the total sample), do not present any signs of X-ray absorption ($\rm N_H<10^{21.5}\,cm^{-2}$). The quality of the SED fitting of these systems is good, based on the $\chi ^2_{\rm red}$ values ($\chi ^2_{\rm red} \lessapprox 2$). SED analysis reveals that the AGN emission is obscured in the optical wavelengths. The optical/mid-IR criteria of \cite{Yan2013}, do not classify these AGN as optically red sources. Furthermore, their optical spectra present broad lines. Thus, there are no indications to corroborate with X-CIGALE that these AGN are absorbed. Therefore, we do not find a plausible explanation for these outliers. However, they are only $<0.4\%$ of our sample ($\sim 1\%$ if we lower the threshold of the $\rm L_X$ difference to 0.3\,dex).

On the right panel of Fig. \ref{lx_comp}, we plot the X-ray luminosity calculations of X-CIGALE when the X-ray flux is not included in the SED vs. the $\rm L_X$ from the input catalogue. In this case, the scatter is significantly larger compared to the left panel while the average error of the $\rm L_X$ estimates from X-CIGALE, increases by $\approx 4\times$ (green lines). When there is no f$_X$ in the SED, the $\rm L_X$ estimates should not be taken at face value. X-CIGALE uses the AGN module (SKIRTOR) to output plausible $\rm L_{2500\,\AA}$ values. Since there is no X-ray information to further constrain the $\rm L_{2500\,\AA}$ parameter by connecting it to the observed X-ray flux, the algorithm provides $\rm L_X$ estimations, using the full range of $\rm \alpha _{ox}$ (allowed by the Just et al. relation and the $\rm \alpha _{ox}$ dispersion) and weighs over these possible values. Thus, although the code provides $\rm L_X$ estimates, these values should be taken with caution.

\subsection{The efficiency of X-CIGALE to connect the X-ray - UV luminosity}

As mentioned, $\rm \alpha _{ox}$, $\rm L_{2500 \AA}$ and $\rm L_{2\,keV}$ are the three parameters that are important for X-CIGALE to connect the X-rays with the UV and thus the other wavelengths during the SED fitting process. In section \ref{appendix_alpha_l2500}, we assess the efficiency of X-CIGALE to constrain these three parameters, by using the mock analysis (see section \ref{sec_mock}). In this section, we compare the algorithm's calculations of $\rm L_{2\,keV}$ and $\rm L_{2500 \AA}$.

Fig. \ref{just_relation}, compares the $\rm L_{2\,keV}$ and $\rm L_{2500\,\AA}$ luminosities with the observed relations of \cite{Just2007} (blue line), used by the code as input (see section \ref{sec_sed_analysis})  and \cite{Lusso2016} (green line). Grey line presents the fit on our measurements. There is an overdensity of sources above the Just et al. relation (at high luminosities). This is due to selection bias. XMM-XXL has a low exposure time  and therefore our X-ray sample is biased towards high luminosity sources.

\subsection{The effect of the extinction law and temperature of polar dust}
\label{sec_extinction_curve} 

In section \ref{sec_agn_emission}, we mentioned that for the polar dust estimation an SMC extinction curve is adopted and the grey-body  dust temperature is set to $100\,K$. The effect of the addition of polar dust in the SED fitting will be discussed in Section \ref{ebv_effect}. In this section, we examine whether the adoption of different extinction curves and dust temperatures affect the polar dust contribution (through $\rm E_{B-V}$) and the AGN fraction calculations. 

Apart from the SMC extinction curve, X-CIGALE includes the choice of the empirical extinction curves of \cite{Calzetti2000} and \cite{Gaskell2004}. Figures \ref{fig_ebv_smc_vs_calzetti} and  \ref{fig_ebv_smc_vs_gaskell} present the difference of the polar dust estimations between the SMC and the \cite{Calzetti2000} and the SMC and \cite{Gaskell2004} curves, respectively. Both distributions are highly peaked and therefore the choice of the extinction curve does not affect the polar dust measurements. In Figures \ref{fig_frac_smc_vs_calzetti} and \ref{fig_frac_smc_vs_gaskell}, we repeat the same test for the AGN fraction values. Again, the adoption of different extinction curves does not affect the $\rm frac_{AGN}$.    

In Figures \ref{fig_ebv_100_vs_75}, \ref{fig_ebv_100_vs_50} and Figures \ref{fig_frac_100_vs_75}, \ref{fig_frac_100_vs_50}, we test whether different temperatures for the grey-body dust re-emission affect the polar dust and AGN fraction calculations, respectively. Specifically, we plot the difference of the aforementioned parameters using values of $100\rm \,K$ (the value used throughout our analysis) as well as $75\rm \,K$ and $50\rm \,K$. All distributions are highly peaked at zero. We obtain similar results when we increase the temperature to $200\,\rm K$. We conclude that the choice of the parameters defining the grey-body re-emission does not affect our results.

\subsection{The effect of {\it Herschel} photometry}

Far-IR data combined with mid-IR photometric bands are known to improve the star-formation rate (SFR) estimations of galaxies hosting an AGN \citep{Hatziminaoglou2009, Stanley2018, Masoura2018} since they constrain more efficiently the AGN contribution to the IR luminosity of the host galaxy. \cite{Masoura2018} used 608 X-ray AGN in XXL with $\it Herschel$ detection and found that SFRs estimated without far-IR photometry are systematically underestimated compared to SFRs with {\it Herschel} (see their Fig. 4). In this section, we examine the effect of {\it Herschel} photometry on our estimations. Specifically, we wish to test whether the absence of {\it Herschel} photometry affects the AGN fraction and polar dust estimates of X-CIGALE.

For this part of our analysis, we restrict our sample to only those AGN with reliable ($\rm SNR>3$) SPIRE photometry (in addition to the criteria mentioned in section \ref{sec_finalsample}). This will maximise any likely effect on the SED fits. This reduces our X-ray dataset to 328 AGN. We run X-CIGALE twice: one time using {\it Herschel} photometry and a second time without (both runs use the X-ray flux of the sources in the fitting process).

Figures \ref{fig_ebv_herschel} and \ref{fig_frac_herschel} present the difference of the polar dust and AGN fraction distributions, from the two runs, respectively. We notice that the addition of {\it Herschel} photometry slightly reduces the AGN fraction measurements. Although, the distribution of the difference of the AGN fraction estimations peaks at zero, there is a tail at negative values. Specifically, there are 73 sources ($\approx 22\%$ of this sample) for which the addition of {\it Herschel} photometry reduces their AGN fraction value by more than 0.15 (which is equivalent to the error of the two $\rm frac_ {AGN}$ estimations added in quadrature, see next). This is also quantified by the mean $\rm frac_ {AGN}$ values, $\rm frac_{AGN, {\it Herschel}}=0.36\pm0.08$ and $\rm frac_{AGN, no\,{\it Herschel}}=0.43\pm0.12$. We also notice that including far-IR photometry increases the statistical significance of the estimations from $3.6\,\sigma$ to $4.5\,\sigma$ (the significance is defined as the bayesian value over the error). The addition of {\it Herschel} photometry does not affect polar dust measurements. We also examine whether the addition of {\it Herschel} photometry affects other parameters estimated by the SED fitting and specifically the X-ray luminosity and the $\rm L_{2500 \AA}$. The distribution of the difference of the $\rm L_X$ and $\rm L_{2500 \AA}$ parameters with and without far-IR photometry is presented in Figures \ref{lx_herschel} and \ref{l2500_herschel}. Both distributions are highly peaked at zero with small tails at both sides. Thus, inclusion of {\it Herschel} photometry does not seem to affect the estimations of these two parameters.


\begin{figure}
\centering
\begin{subfigure}{.27\textwidth}
  \centering
  \includegraphics[width=1.\linewidth]{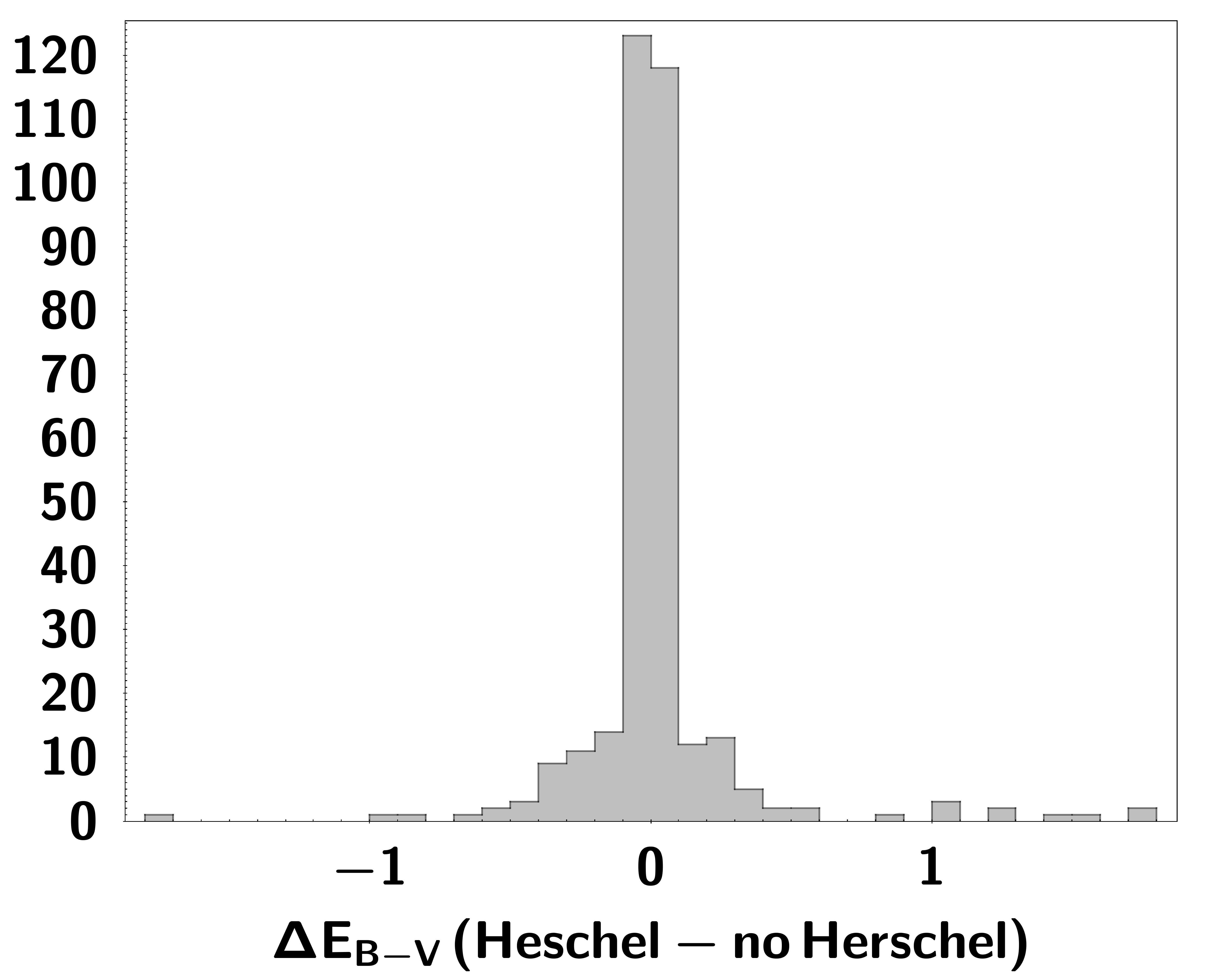}
  \caption{}
  \label{fig_ebv_herschel}
\end{subfigure}
\begin{subfigure}{.27\textwidth}
  \centering
  \includegraphics[width=1.\linewidth]{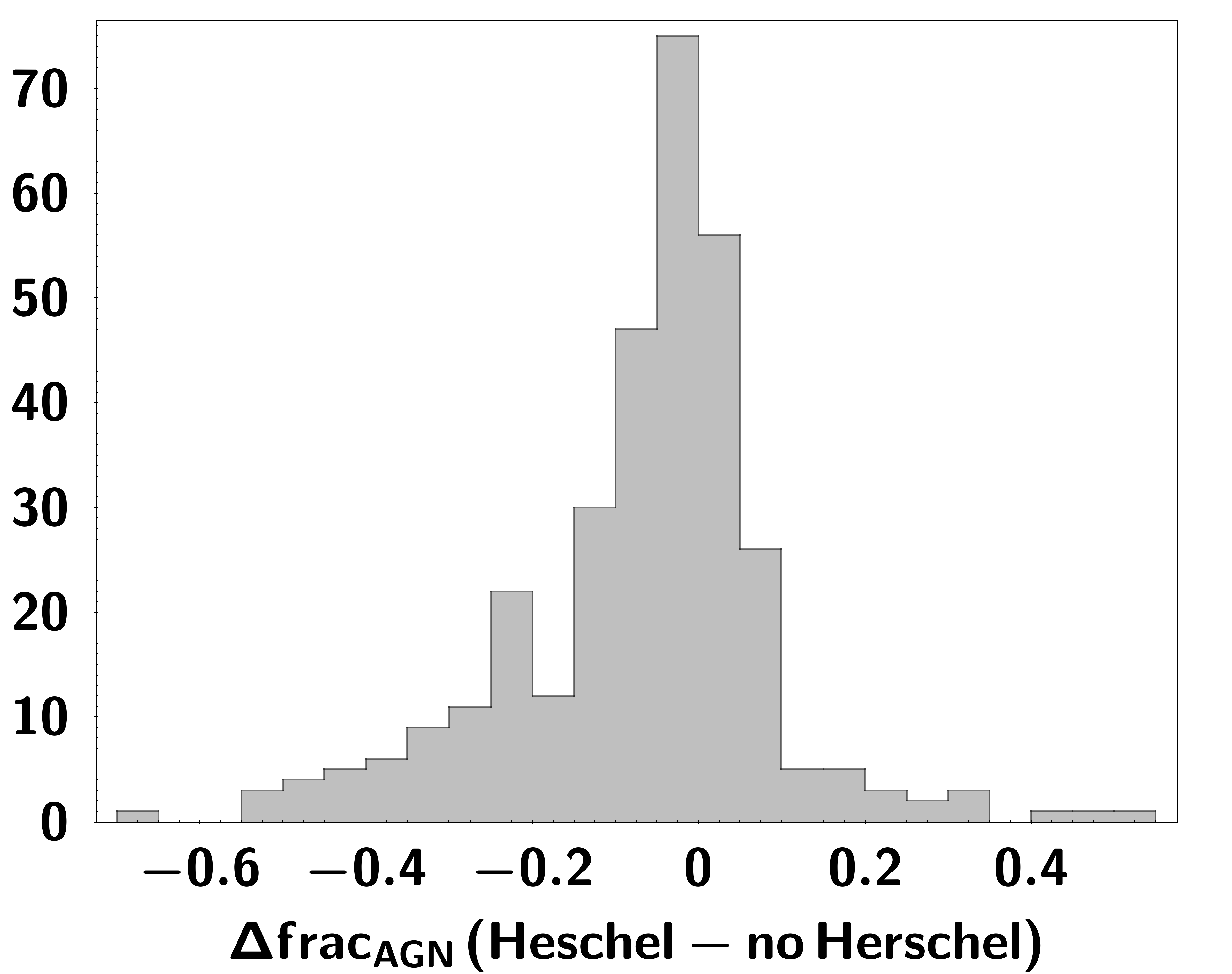}
  \caption{}
  \label{fig_frac_herschel}
\end{subfigure}
\begin{subfigure}{.27\textwidth}
  \centering
  \includegraphics[width=1.\linewidth]{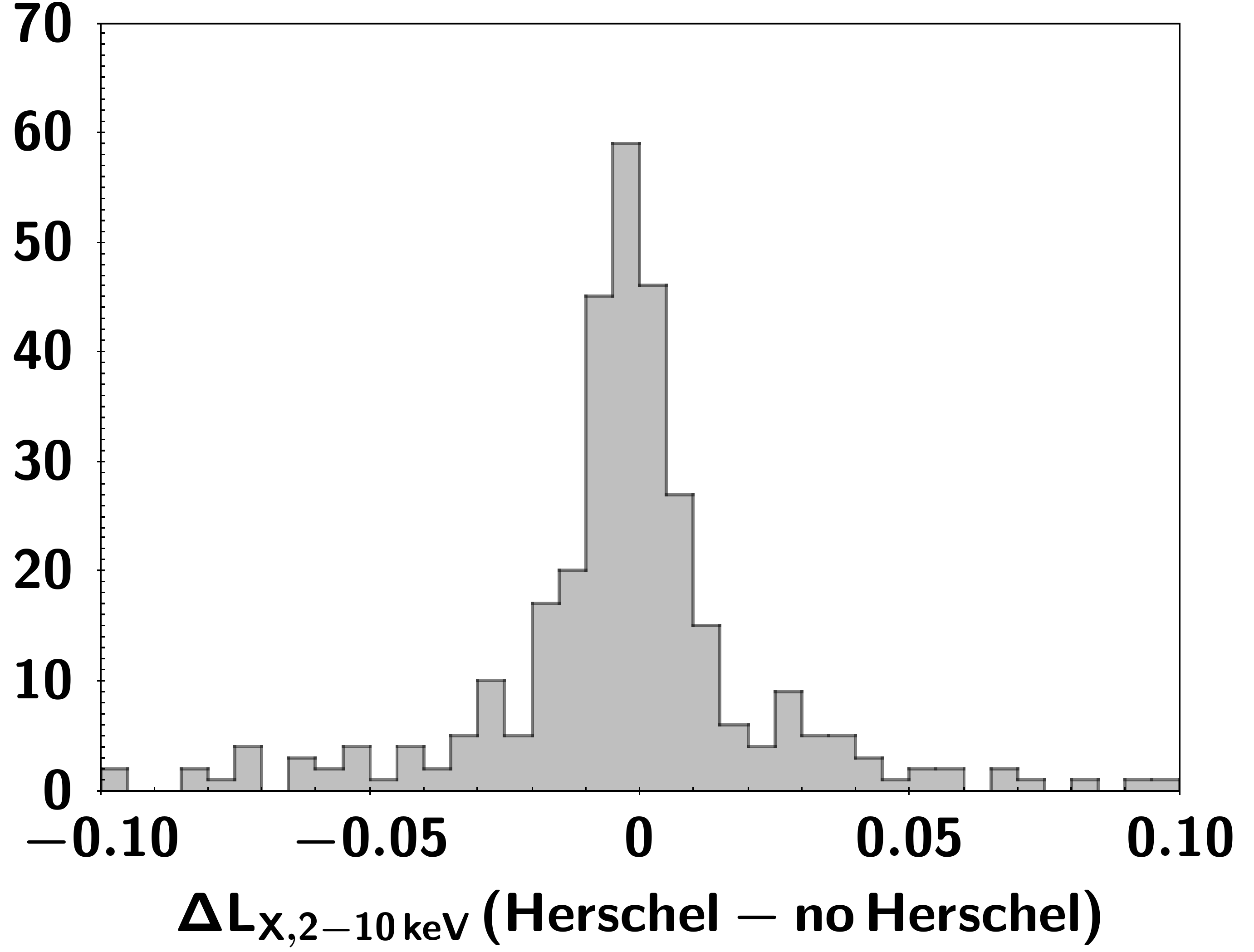}
  \caption{}
  \label{lx_herschel}
\end{subfigure}
\begin{subfigure}{.27\textwidth}
  \centering
  \includegraphics[width=1.\linewidth]{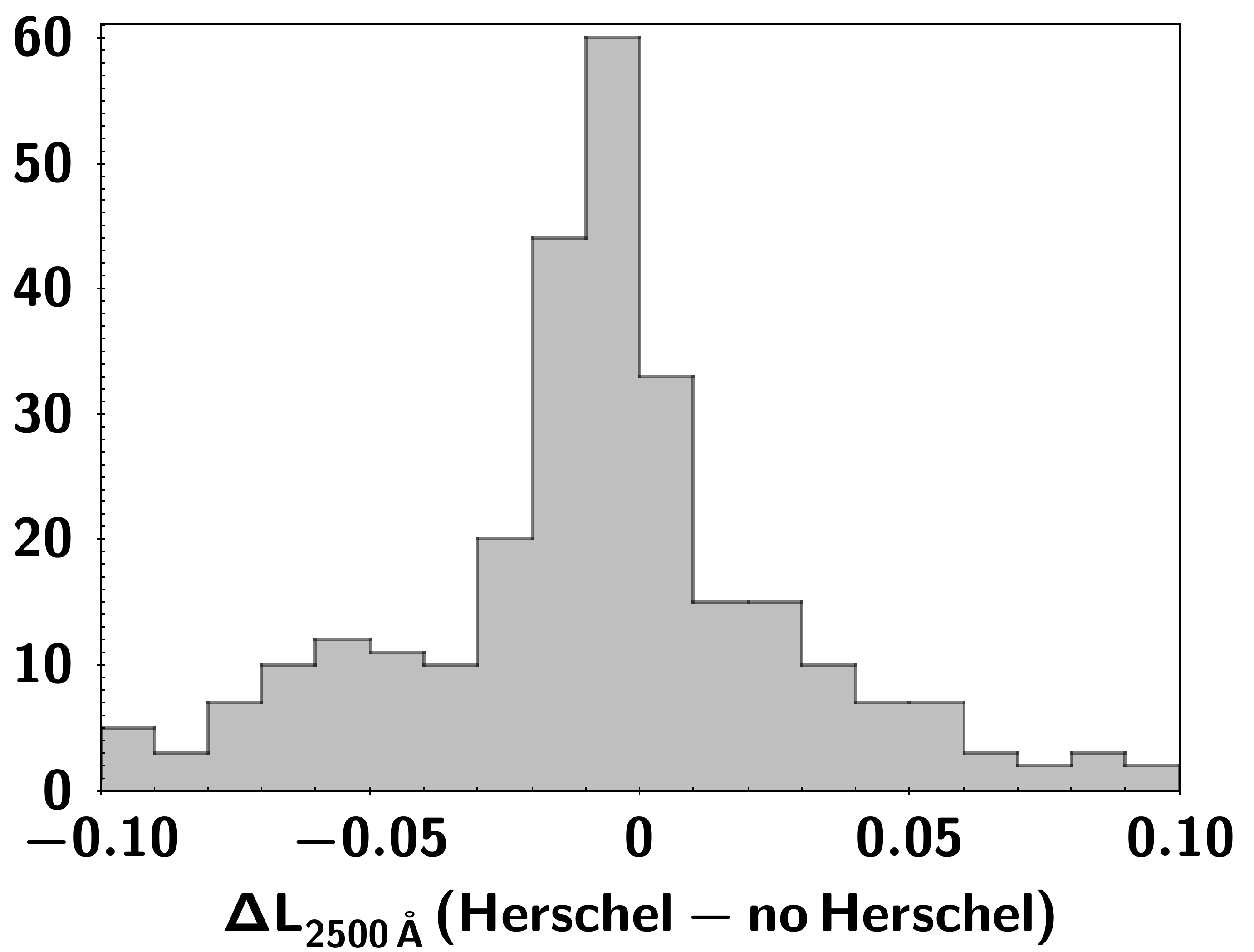}
  \caption{}
  \label{l2500_herschel}
\end{subfigure}
\caption{Distribution of the difference of polar dust, AGN fraction and X-ray and UV luminosity measurements, with and without $\it Herschel$ photometry. Polar dust and luminosity calculations are not affected by the existence of far-IR photometry. However, addition of far-IR photometry reduces the $\rm frac _{AGN}$. This is quantified by the mean values, $\rm frac_{AGN, Herschel}=0.36\pm0.08$ and $\rm frac_{AGN, no\,Herschel}=0.43\pm0.12$.} 
\label{herchel_comp}
\end{figure}

\begin{figure*}
\centering
\begin{subfigure}{.5\textwidth}
  \centering
  \includegraphics[width=1.\linewidth]{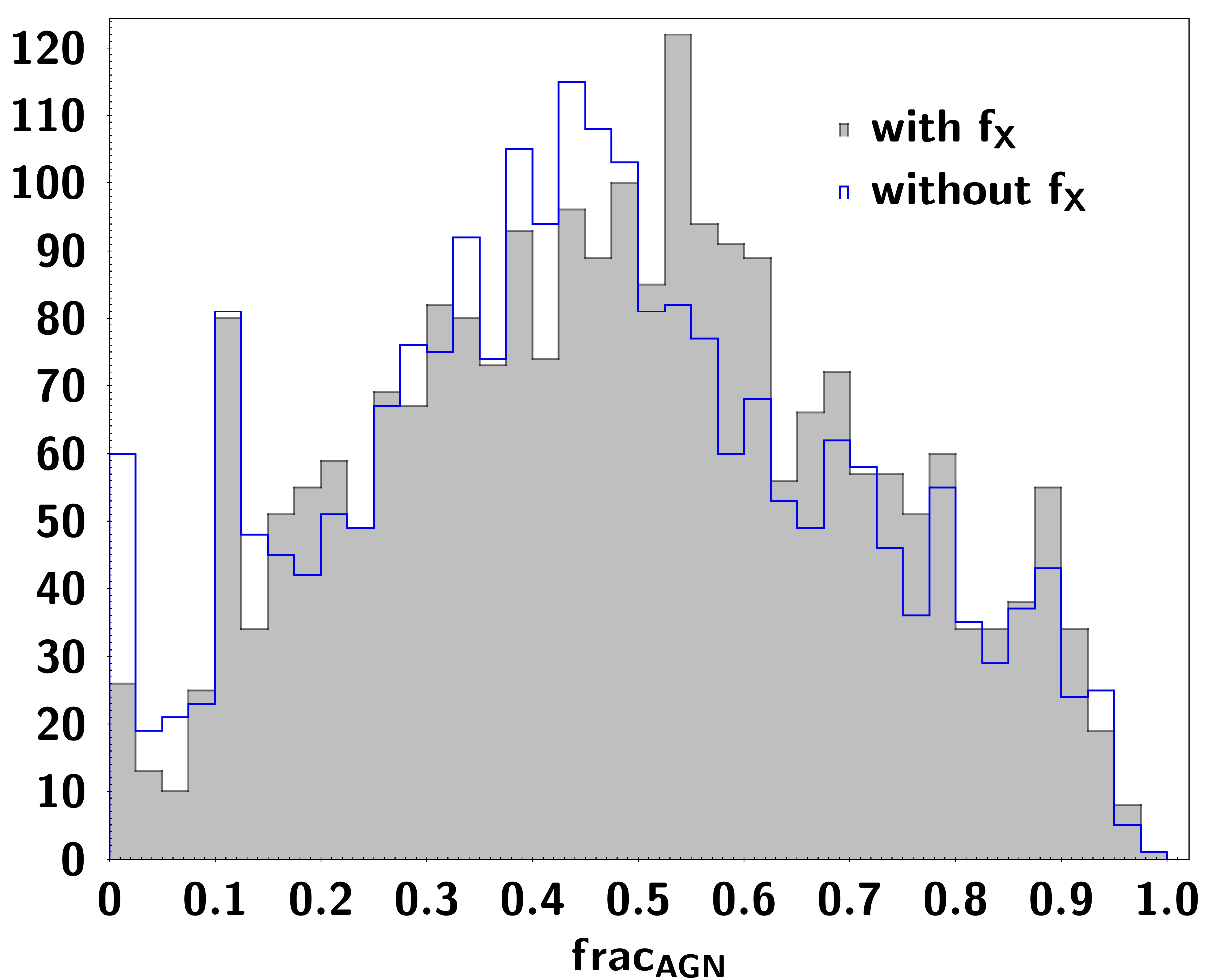}
  \label{frac_distrib}
\end{subfigure}%
\begin{subfigure}{.5\textwidth}
  \centering
  \includegraphics[width=1.\linewidth]{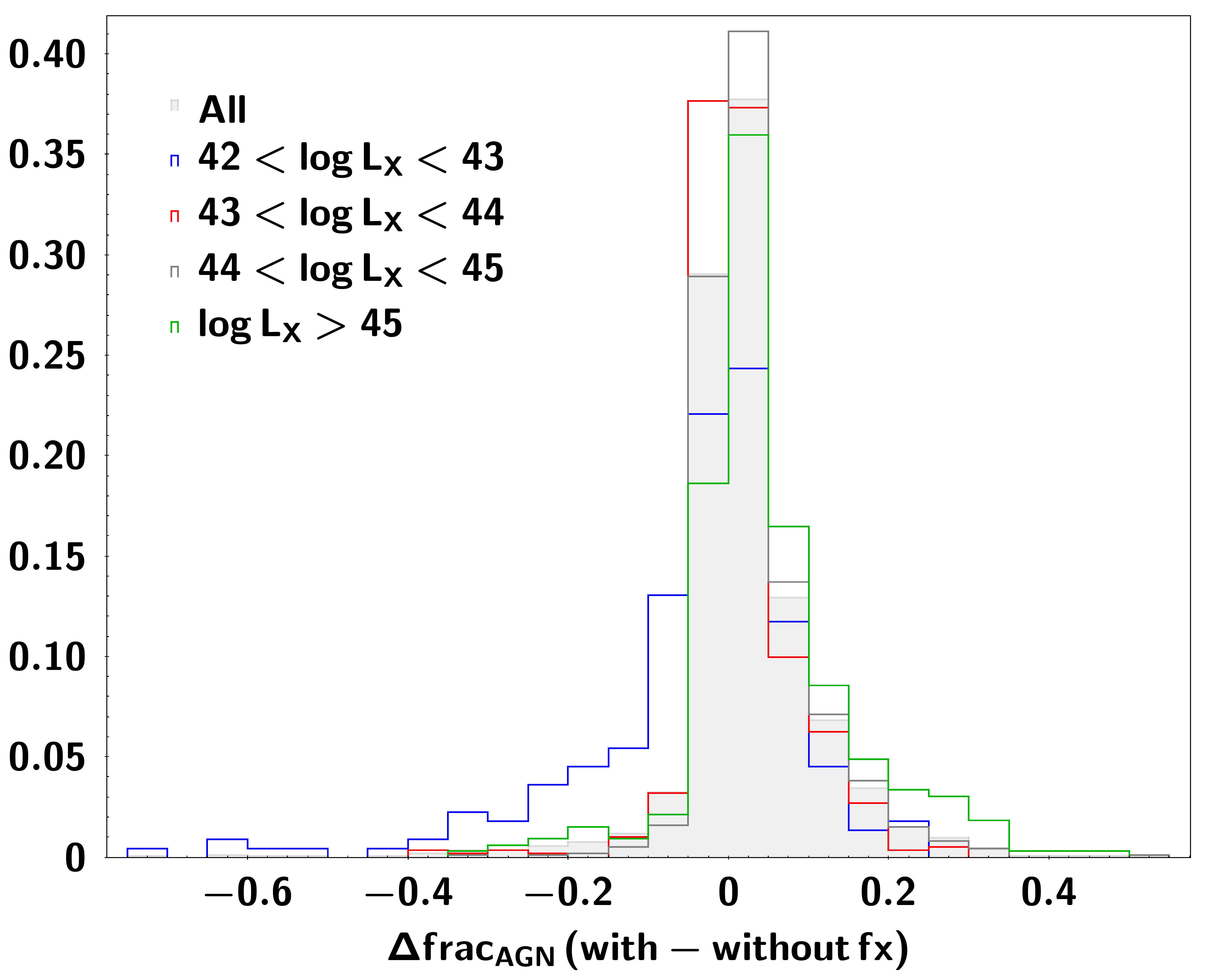}
  \label{frac_diff}
\end{subfigure}
\caption{Left: distribution of the AGN fraction, with X-ray flux (black shaded area) and without X-ray flux (blue histogram).  Right: distribution of the difference of the AGN fraction, with and without X-ray flux, for different luminosity bins. All distributions are normalised to unity. A tail appears at positive values, i.e., the AGN fractions are higher when X-ray flux is included in the SED fitting, for AGN with $\rm {L_{X}>10^{43}\, erg\,s^{-1}}$ that becomes more prominent for the most luminous sources ($\rm {L_{X}>10^{45}\, erg\,s^{-1}}$). On the other hand, there is a tail at negative values, i.e. the AGN fractions are lower with X-ray flux, for the fainter AGN ($\rm {L_{X}<10^{43}\, erg\,s^{-1}}$, blue line).}
\label{agn_fraction}
\end{figure*}

\begin{figure}
\centering
  \centering
  \includegraphics[width=1.\linewidth]{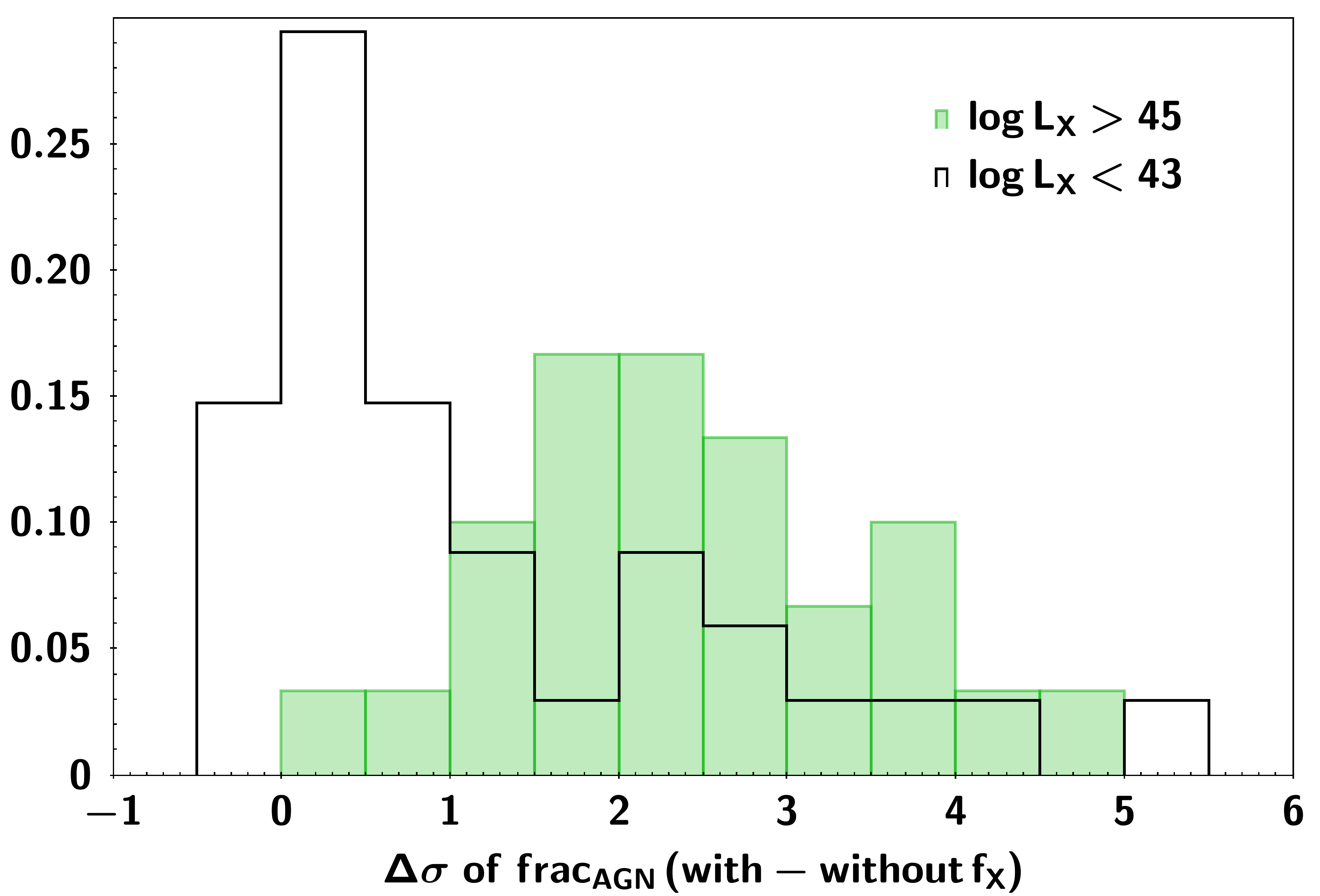}
	\caption{The difference of the significance, $\Delta \sigma$, of the AGN fraction estimations with and without X-ray information, for sources with $\rm {L_{X}>10^{45}\, erg\,s^{-1}}$ and difference in $\rm frac_{AGN}>0.2$ (shaded histogram) and sources with $\rm {L_{X}<10^{43}\, erg\,s^{-1}}$ and difference in  $\rm frac_{AGN}<-0.2$. The statistical significance of the AGN fraction measurements improves significantly with the inclusion of the X-ray flux, especially for the most luminous AGN.}
\label{frac_significance}
\end{figure}

\begin{figure*}
\centering
\begin{subfigure}{.5\textwidth}
  \centering
  \includegraphics[width=1.\linewidth]{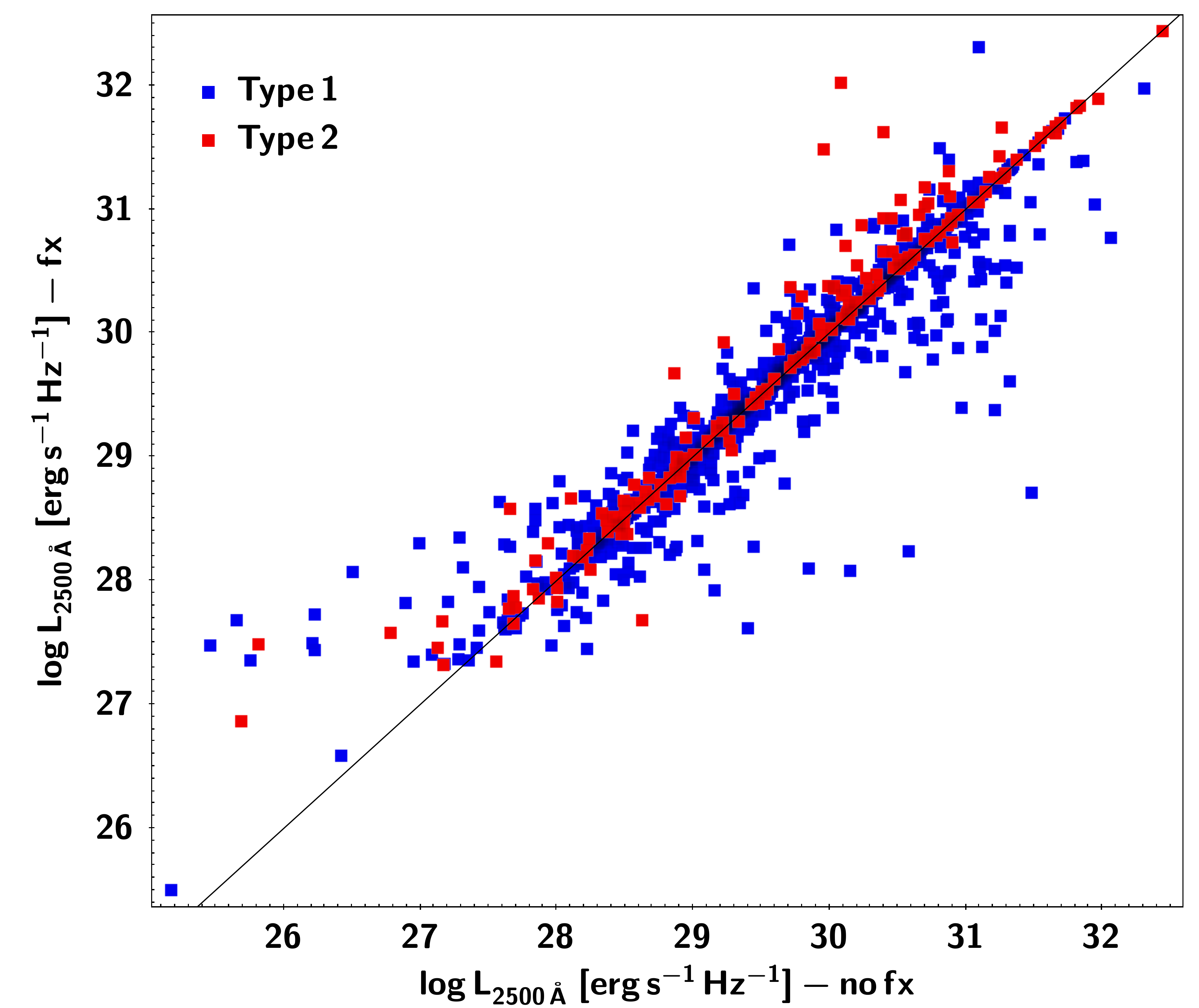}
  \label{}
\end{subfigure}%
\begin{subfigure}{.5\textwidth}
  \centering
  \includegraphics[width=1.\linewidth]{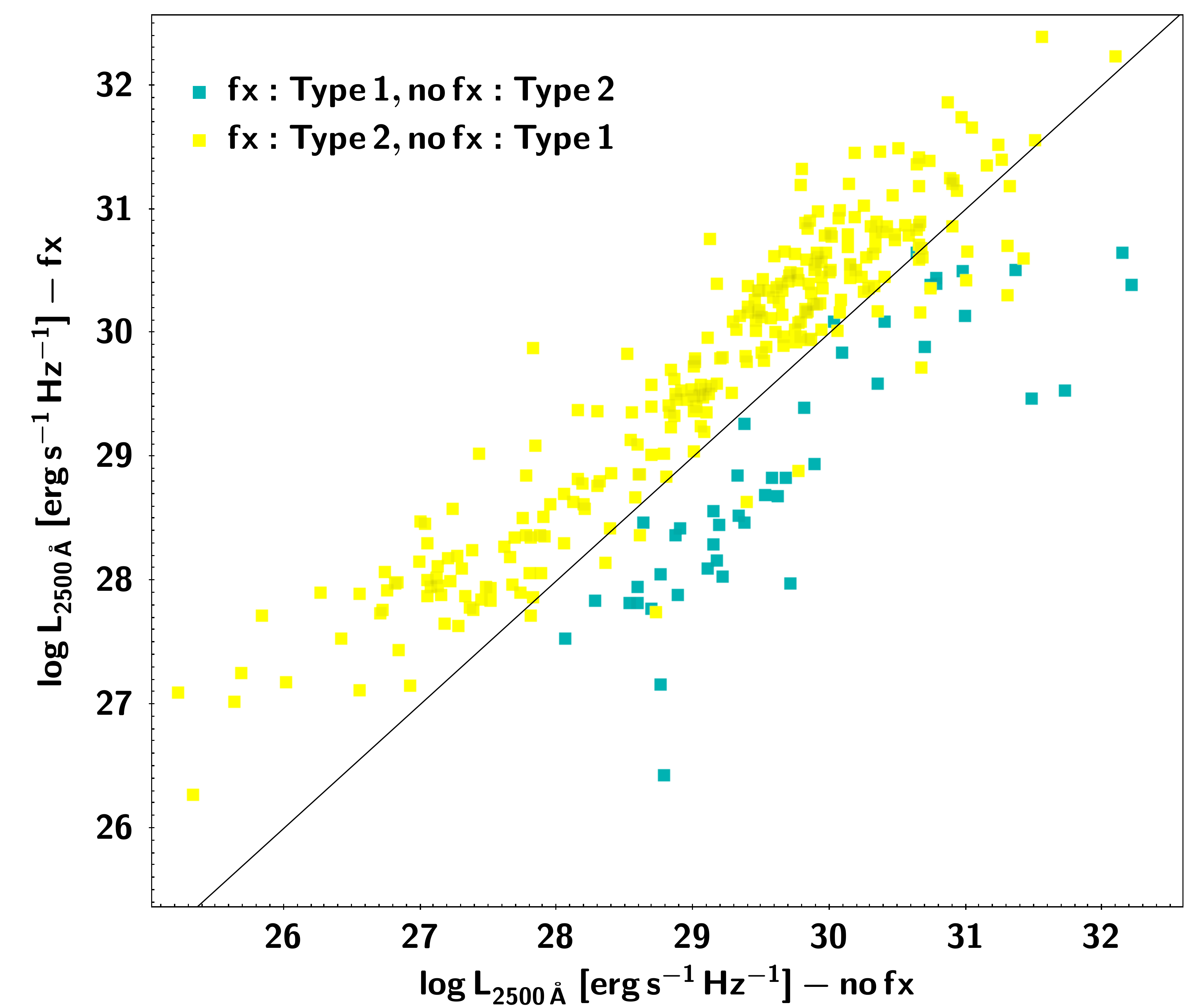}
  \label{}
\end{subfigure}
\caption{Comparison of the $\rm L_{2500\,\AA}$ values with and without X-ray flux in the fitting process. Left panel: Comparison for sources that have classified with the same type, with and without X-ray flux in the SED fitting process. Blue symbols present the results for type 1 sources ($\rm i=30^{\circ}$) and red symbols for type 2 sources ($\rm i=70^{\circ}$). Right panel: Comparison of the $\rm L_{2500\,\AA}$ values for sources with different types from the two runs (with and without X-ray flux). X-CIGALE will decrease/increase the intrinsic accretion power (and thus the intrinsic $\rm L_{2500\,\AA}$) depending on whether the source is type 1/2.}
\label{L2500_comp_fx_nofx}
\end{figure*}

\section{Advantages of X-ray flux and polar dust components}
\label{sec_advantages}

In this section, we examine the advantages that the introduction of the two new features of X-CIGALE, i.e. the  X-ray flux and polar dust, bring in the SED fitting process.

\subsection{The effect of the X-ray flux on the AGN fraction measurements}

One of the strengths of SED decomposition is that it allows disentangling AGN emission from that of the host galaxy emission. This is a crucial issue since the reliability of any further analysis depends on how reliable the SED fitting code can perform this task. The goal of this part of our analysis is to examine if the introduction of the  X-ray information in the SED, allows X-CIGALE to improve its efficiency in estimating robust AGN fractions.

In section \ref{appendix_agnfrac}, we assess the accuracy of X-CIGALE in the estimation of the AGN fraction. Here, we study how the addition of the the X-ray flux in the fitting process affects the AGN fraction measurements. In Fig. \ref{agn_fraction} (left panel), we plot the distribution of the AGN fraction with (black shaded area) and without (blue histogram) the X-ray flux (polar dust is included in both runs). The mean $\rm frac_{AGN}$ value increases from $0.46\pm0.16$ to $0.49\pm0.14$ with the addition of $\rm f_X$. To further examine the effect of the X-ray flux in the AGN fraction estimates, in the right panel we plot the difference of the AGN fraction with and without X-ray flux, for different luminosity bins. The distributions peak at zero, in all cases. However, at high X-ray luminosities ($\rm {L_{X}>10^{43}\, erg\,s^{-1}}$) a tail starts to appear at positive values, i.e., the AGN fraction tends to be higher when the X-ray flux is included in the fitting process. This tail becomes more prominent at the highest luminosity bin. Specifically, for AGN with $\rm {L_{X}>10^{45}\, erg\,s^{-1}}$ the mean AGN fraction increases from $0.58\pm0.16$ without X-ray flux to $0.65\pm0.13$ with X-ray flux. The opposite trend is present for low luminosity AGN ($\rm 10^{42}<{L_{X}<10^{43}\, erg\,s^{-1}}$). In this case, the mean $\rm frac_{AGN}$ reduces from $0.28\pm0.13$ to $0.23\pm0.09$ with the X-ray flux. 

The statistical significance of the AGN fraction measurements improves with the inclusion of the X-ray flux, as can be seen by the numbers presented above. This is also illustrated in Fig. \ref{frac_significance}, where we plot the difference of the significance, $\Delta \sigma$, of the AGN fraction estimations with and without X-ray information, for sources with $\rm {L_{X}>10^{45}\, erg\,s^{-1}}$ and difference in $\rm frac_{AGN}>0.2$ (shaded histogram) and sources with $\rm {L_{X}<10^{43}\, erg\,s^{-1}}$ and difference in  $\rm frac_{AGN}<-0.2$. Thus, we conclude that the addition of the X-ray flux in the SED fitting results in more robust AGN fraction measurements while its effect on $\rm frac_{AGN}$ is small and depends on the AGN power.

\begin{figure}
\centering
  \centering
  \includegraphics[width=1.\linewidth]{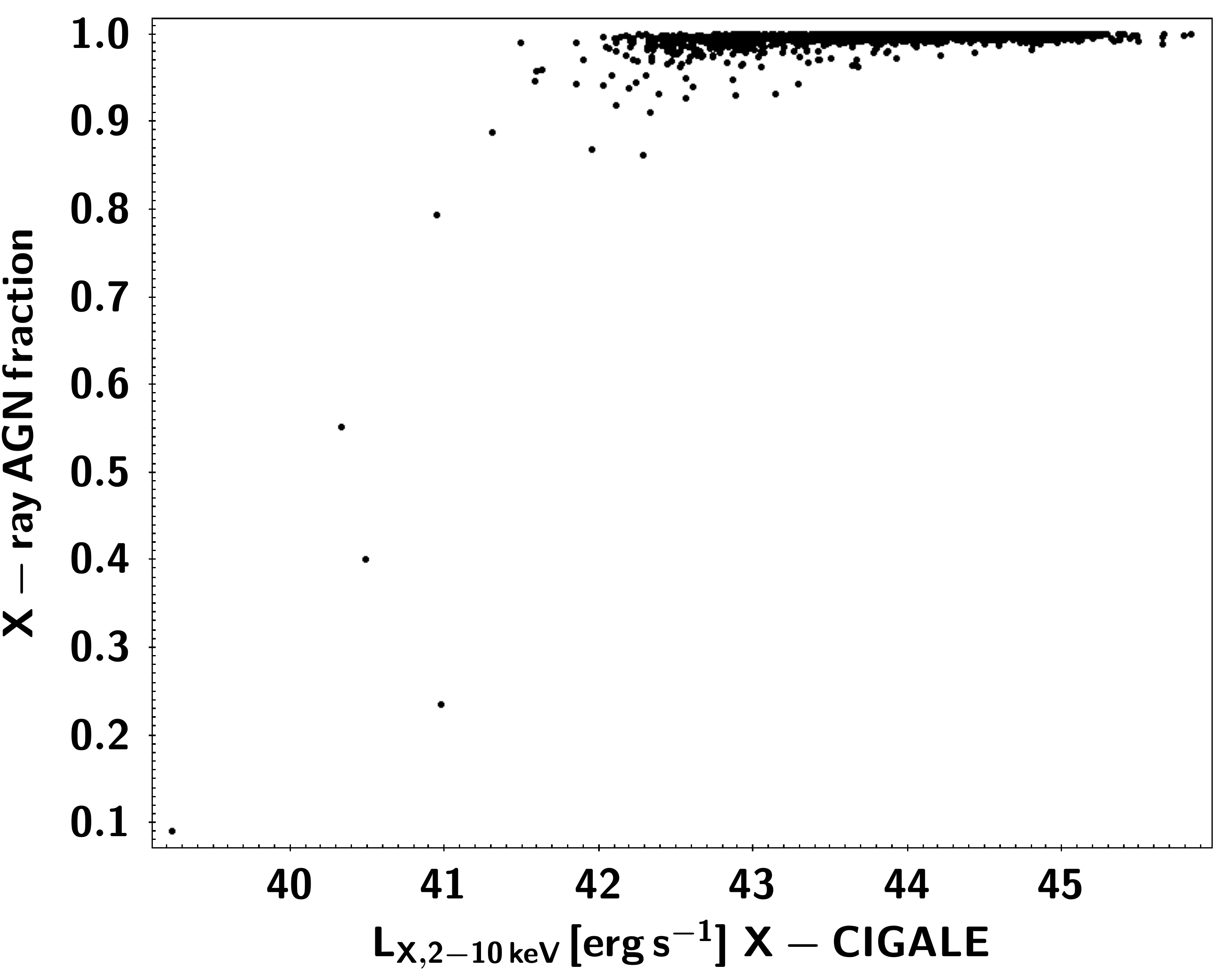}
	\caption{X-ray AGN fraction vs. X-ray luminosity, estimated by X-CIGALE. The X-ray AGN fraction is defined as the ratio of the X-ray AGN emission to the total X-ray emission of the galaxy ($\rm AGN+binaries+hot gas$). As expected, for the vast majority of sources the AGN X-ray emission contributes more than 90\% of the total X-ray emission of the galaxy.}
\label{fig_xray_fraction}
\end{figure}

\subsection{The effect of X-ray flux on the estimation of the $\rm L_{2500\,\AA}$ and AGN type}

Another important parameter of the SED fitting is the intrinsic $\rm L_{2500\,\AA}$ luminosity. Additionally to the reasons mentioned in section \ref{sec_sed_analysis}, $\rm L_{2500\,\AA}$ is also a proxy of the AGN power. Here, we study the effect of the X-ray flux on the estimation of $\rm L_{2500\,\AA}$. Fig. \ref{L2500_comp_fx_nofx} (left panel) compares the $\rm L_{2500\,\AA}$ estimations with and without X-ray flux in the SED, for type 1 and 2 sources. The classification is based on their inclination angle values, as calculated by X-CIGALE. 2264 sources satisfy our selection criteria, described in section \ref{sec_finalsample}, in both runs (with and without X-ray flux). In Fig. \ref{L2500_comp_fx_nofx} we have considered only the 1941 sources (86\%) for which the inclination angle from the best fit  has the same value in the two runs. For both types, we observe a scatter, but no systematic effect between the two estimations.

1810/2264 AGN are classified as type 1 with the X-ray flux, but 47 ($\sim 2.5\%$) become type 2 without the X-ray flux. On the other hand, from the 454/2264 sources classified as type 2 with $\rm f_X$, 276 ($\sim 60\%$) change to type 1 without X-ray flux. The 323 (47+276) sources with different classification in the two runs, are shown in the right panel of Fig.\,\ref{L2500_comp_fx_nofx}. X-CIGALE increases $\rm L_{2500\,\AA}$  when the type changes from type 1 to type 2 with the inclusion of the X-ray flux and lowers $\rm L_{2500\,\AA}$ when the type changes from type 2 to type 1 with $\rm f_X$. This result is expected, considering the way that X-CIGALE works. (mid-) IR emission is considered anisotropic, i.e., it depends on the viewing angle. The same source will have lower IR flux when viewed edge on (type 2) compared to face on \citep[see Fig. 4 in][]{Stalevski2012}. On the other hand, X-ray flux is considered isotropic. Therefore, a type 2 source will have a higher $\rm {\frac{L_{X,intrinsic}}{L_{IR}}}$ (and $\rm {\frac{L_{2500\,\AA,intrinsic}}{L_{IR}}}$) than a type 1 source. For a given observe IR emission, X-CIGALE will decrease/increase the intrinsic accretion power (and thus the intrinsic $\rm L_{2500\,\AA}$) depending on whether the source is type 1/2.

Our analysis revealed that the inclusion of the X-ray information does not significantly affect the classification of sources that are type 1 based on the run with $\rm f_X$, but it does change the characterisation of the majority of the AGN classified as type 2 with X-ray flux. To investigate further, the different classification for the 276 AGN classified as type 2 when the X-ray information is included in the SED, but classified as type 1 without f$_X$, we compare their X-CIGALE classification with that from optical (SDSS) spectra \citep{Menzel2016}. Among those sources that lie at $\rm z<1$, $\sim 85\%$ are classified as Narrow Line AGN \citep[NLAGN2; see section 3.3.2 in][]{Menzel2016},   i.e. the optical spectrum agrees with the classification of X-CIGALE when using the X-ray flux. On the other hand, at $\rm z>1.5$, $\sim 85\%$ of the AGN present broad lines in their optical spectra (BLAGN1), i.e. their classification does not agree with that of X-CIGALE when the X-ray flux is included. We conclude that, at $\rm z<1$, the addition of the X-ray information increases the reliability of the SED fitting regarding the classification type of a source. At high redshifts, though, X-CIGALE seems to misclassify some sources: among the 703 AGN at $\rm z>1.5$ in our total sample, 103 ($\approx 15\%$) are classified as type 2 with X-ray flux and as type 1 without.

We further investigate plausible reasons for the misclassification of sources at high redshift, when $\rm f_X$ is added in the fitting process. The vast majority of these sources lack photometry above 5 microns. In the absence of IR photometry and without considering the X-ray flux, the introduction of a type 1 component gives a flexibility to fit the UV/optical data while a type 2 component does not contribute to this emission. Thus, without $\rm f_X$, X-CIGALE is more likely to classify high redshift sources as type 1, with the risk to over-fit the UV/optical data. Inclusion of X-ray flux, provides an additional constraint for a type 1 template (via the Just et al. relation), whereas the code can freely scale a type 2 template to match any given X-ray flux. Thus, with $\rm f_X$, the code will preferentially classify sources as type 2. In this configuration (SED with only UV/optical data), the difference between a type 1 and a type 2 component is not meaningful.

We conclude, that the misclassification of AGN at high redshifts is mostly due to lack of available IR photometry which makes the type assignation almost unconstrained. Future surveys \citep[{\it {JWST}};][]{Gardner2006} could provide mid-IR photometry for these distant AGN.

\begin{figure}
\centering
  \includegraphics[width=0.9\linewidth]{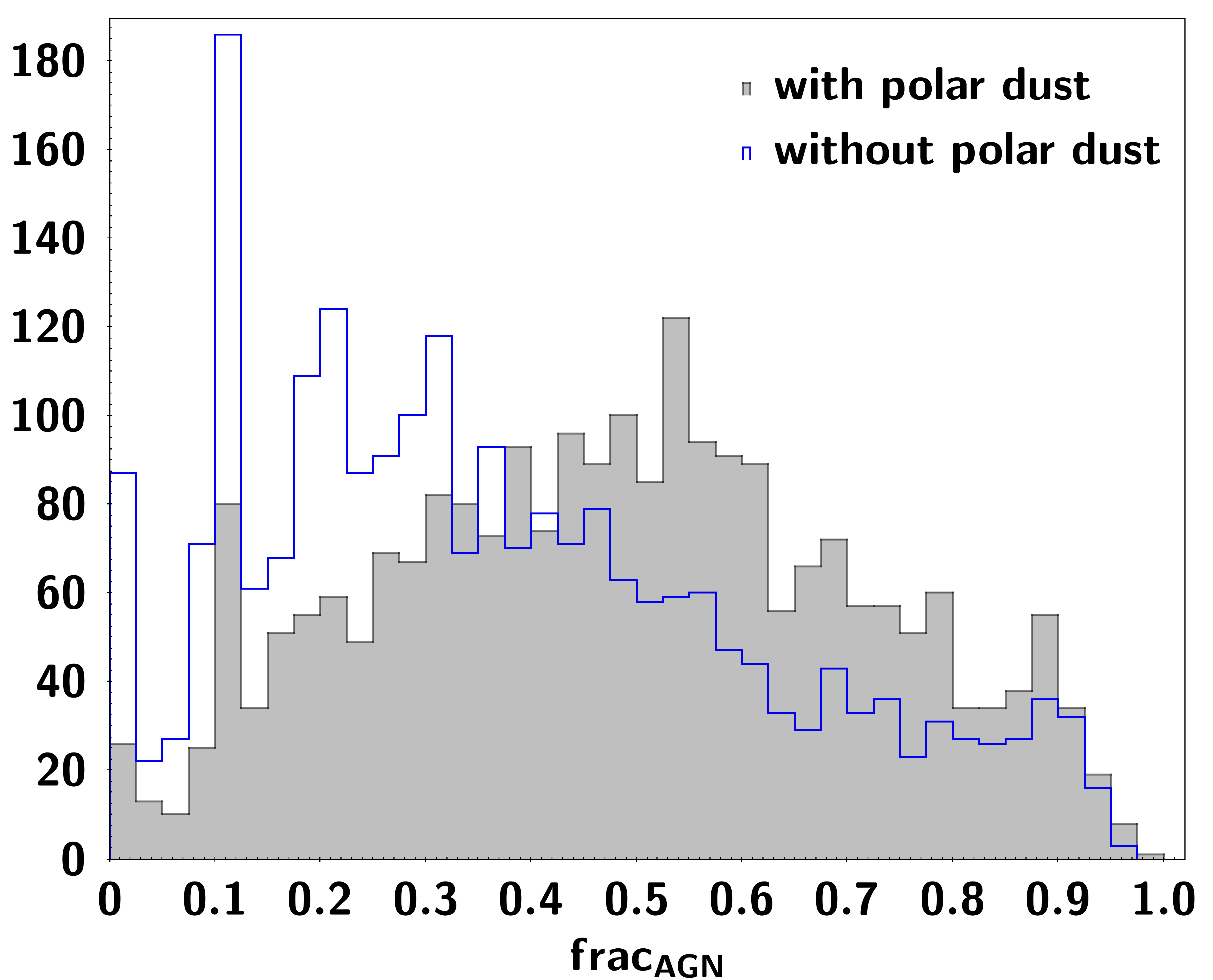}
  \caption{Distribution of the AGN fraction, with polar dust (black shaded area) and without polar dust (blue histogram). The addition of polar dust in the fitting process increases the AGN fraction.}
  \label{agn_fraction_polar}
\end{figure}

\begin{figure*}
\centering
  \includegraphics[width=0.9\linewidth]{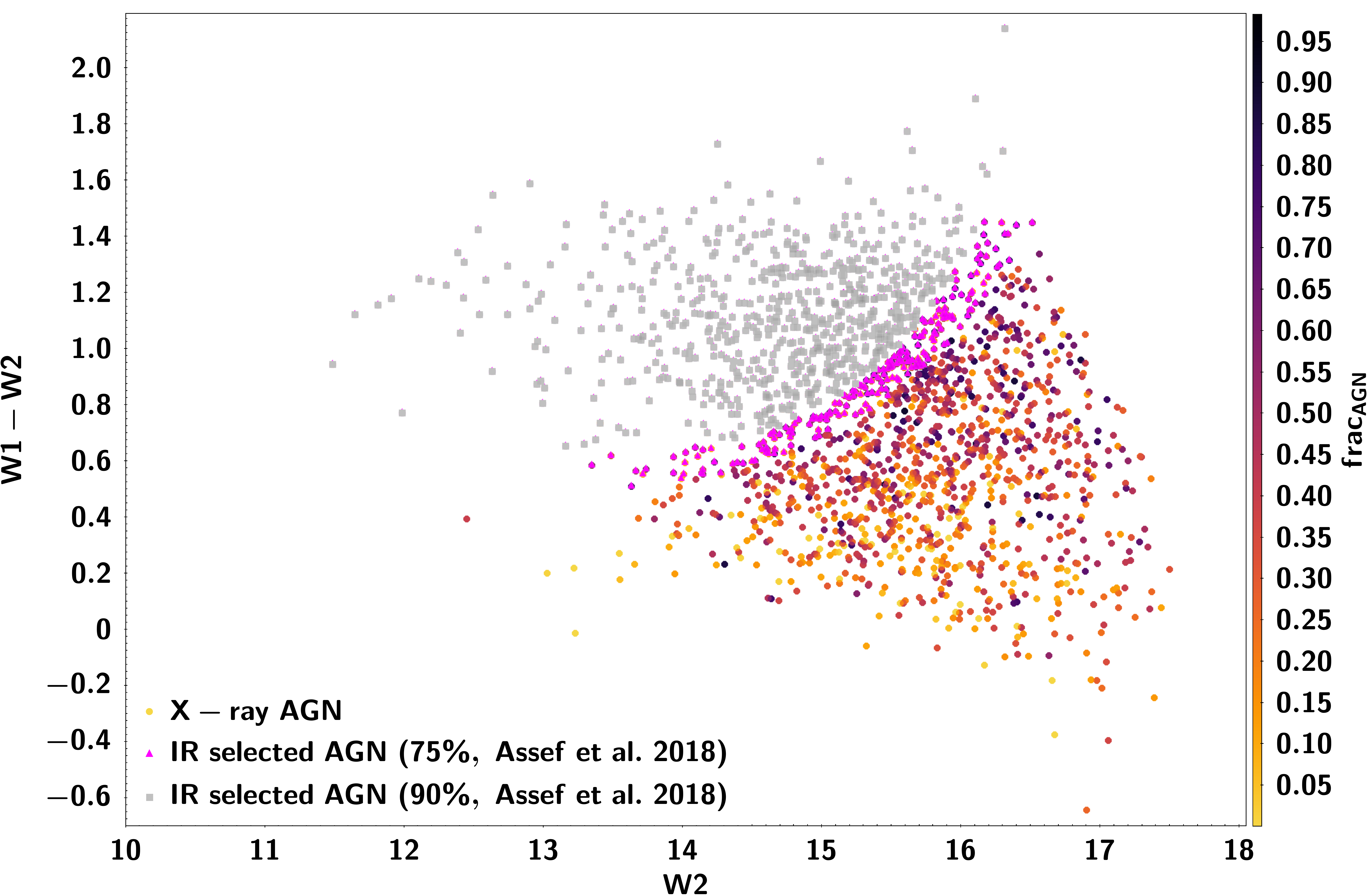}
  \caption{W1-W2 vs. W2 for the X-ray AGN in our sample (circles). Sources selected as IR AGN with 75\% and 90\% confidence, using the criteria of Assef et al. 2018 are in magenta and grey, respectively. The remaining sources of our X-ray AGN sample are presented with circles, colour coded based on the AGN fraction measurements of X-CIGALE. $\sim 97\%$ of IR selected AGN have $\rm frac_{AGN}>0.2$ ($\sim 92\%$ with $\rm frac_{AGN}>0.3$). This percentage drops to $71\%$ ($55\%$ with $\rm frac_{AGN}>0.3$), without polar dust.}
  \label{assef_colour}
\end{figure*}

\begin{table*}
\caption{}
\centering
\setlength{\tabcolsep}{1.5mm}
\begin{tabular}{lcccc}
       \hline
{SED configuration} & {X-ray AGN} & \multicolumn{2}{c}{IR detected AGN} & {non-IR detected AGN} \\
& & {75\% Assef} & {90\% Assef} &  \\
       \hline
& 1956 & {889/1956 (45\%)} & {689/1956 (35\%)} & 1067/1956 (55\%) \\ 
\\
Polar dust ($\rm frac_{AGN}>0.2$)  &  &  851/889 (96\%)   & 671/689 (97\%) & 906/1067 (85\%)  \\
Polar dust ($\rm frac_{AGN}>0.3$)  &  &  795/889 (90\%)   & 631/689 (92\%) & 789/1067 (74\%)  \\  
\\  
w/out polar dust ($\rm frac_{AGN}>0.2$)  &  &  740/889 (83\%)   & 591/689 (86\%) & 757/1067 (71\%)  \\ 
w/out polar dust ($\rm frac_{AGN}>0.3$)  &  &  603/889 (68\%)   & 484/689 (70\%) & 586/1067 (55\%)  \\    
       \hline
\label{table:detection_efficiency}
\end{tabular}
\end{table*}

\begin{figure*}
\centering
  \includegraphics[width=0.9\linewidth]{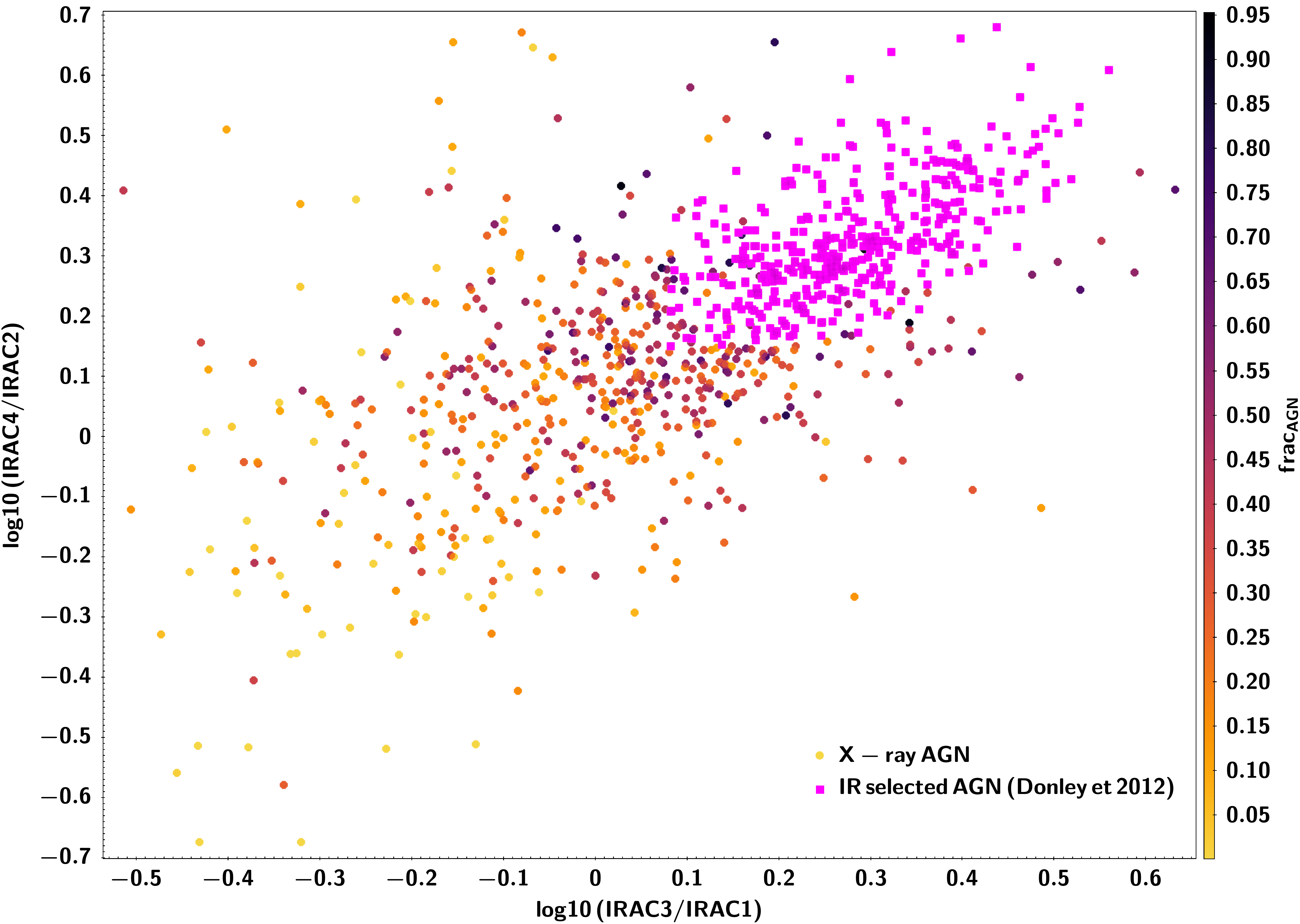}
  \caption{Colour-colour distribution of the X-ray AGN in our sample (circles), using $\it Spitzer$ photometry. Circles are colour coded based on the AGN fraction estimations of X-CIGALE. Sources selected as IR AGN using the colour criteria of Donley et al. 2012, are marked with purple squares.}
  \label{donley_colour}
\end{figure*}

\subsection{Ability to quantify the AGN contribution to the total galaxy X-ray emission}

Star-forming galaxies may emit X-rays that originate from X-ray binaries, supernovae remnants and hot gas \citep[e.g.][]{Fabbiano1989}. Previous studies have shown that X-ray luminosity can be used as a proxy of SFR \citep[e.g.][]{Ranalli2003, Laird2005}. However, in X-ray luminous AGN, like those used in this study, the X-ray emission predominately originates from the supermassive black hole and therefore these systems are usually identified and excluded by studies that use the X-ray emission to trace star formation \citep[e.g.][]{Persic2002, Laird2005}.

As already mentioned, X-CIGALE can also model the X-ray emission that originates from low- and high-mass X-ray binaries and hot gas. Thus, in a similar manner to the $\rm frac_{AGN}$ that quantifies the contribution of the AGN IR emission to the total galaxy IR emission, we define the X-ray AGN fraction as the ratio of the X-ray AGN emission to the total X-ray emission of the galaxy ($\rm AGN+binaries+hot gas$). In Fig. \ref{fig_xray_fraction}, we plot the X-ray AGN fraction as a function of the X-ray luminosity, estimated by X-CIGALE. As expected, for the vast majority of sources the AGN X-ray emission contributes more than 90\% of the total X-ray emission of the galaxy.

Therefore, X-CIGALE by modelling the X-ray emission and its various components offers the capability of estimating the SFR of galaxies not only from IR indicators but also from X-rays, in those systems that are not AGN dominated.

\subsection{Polar dust and AGN fraction}
\label{ebv_effect}

Another important feature of X-CIGALE is polar dust, quantified with $\rm E_ {B-V}$, as a free parameter to account for dust extinction in all AGN, regardless of their classification into type 1 and 2 (see section \ref{sec_agn_emission}). In section \ref{appendix_polar}, we test whether X-CIGALE can successfully constrain $\rm E_ {B-V}$. Here, we study the effect of adding polar dust as a free parameter to the SED fitting results, on the AGN fraction measurements.

Fig. \ref{agn_fraction_polar}, presents the distribution of the AGN fraction with polar dust (black shaded area) and without polar dust (blue histogram). In both runs, X-ray flux is included. The addition of polar dust in the fitting process increases the AGN fraction. The mean $\rm frac_{AGN}$ is $0.37\pm0.12$ without polar dust and $0.49\pm0.14$ with the addition of polar dust. A similar increase is observed when the two runs do not include f$_X$. This trend is also similar regardless of the AGN type, classified based on the inclination angle. The increase of the AGN fraction is expected since the introduction of polar dust allows X-CIGALE to account for obscuring material in the poles of AGN.

\subsection{Polar dust and AGN type}

Repeating the same exercise as for the inclusion of the X-ray flux, we examine whether there are sources for which the classification, based on the estimated inclination angle, changes with the introduction of polar dust ($\rm f_X$ is not included in these runs). Our analysis shows that only $0.01\%$ of the AGN classified as type 2 with polar dust, change classification without it. However, $14\%$ of the sources characterised as type 1 when polar dust is included, change to type 2 without polar dust. Addition of attenuation by polar dust, provides X-CIGALE the flexibility to attribute part of the absorption to polar dust. 

{We compare the classification of those sources that their type changes from type 1 to type 2 when we ignore polar dust, with the classification from optical SDSS spectra. $\sim 70\%$ of the AGN present broad lines in the optical continuum. We conclude that the introduction of polar dust as a free parameter in the SED fitting process, improves the accuracy of X-CIGALE in the classification type of the sources.

\subsection{Polar dust and AGN detection efficiency}
\label{ebv_effect3}
X-ray detection provides the most reliable method to identify AGN. Although, X-rays have proven efficient in penetrating the intervening absorbing material \citep[e.g.][]{Luo2008}, even hard X-rays are absorbed by huge amounts of gas and dust. Thus, X-ray selected AGN are biased against the most heavily absorbed sources. On the other hand, mid-IR surveys are less affected by extinction. The material that absorbs AGN radiation even at hard X-ray energies, is heated by the AGN and re-emits at infrared wavelengths. Therefore, mid-IR selected AGN can detect AGN that are missed by X-ray surveys \citep[e.g.][]{Georgantopoulos2008, Fiore2009}. $\it{Spitzer}$, was the first infrared survey that was used to select AGN via colour selection criteria \citep[e.g.][]{Stern2005, Donley2012}. Later. these techniques were adapted and used for WISE. For instance, \cite{Mateos2012} used a AGN selection method, based on three WISE colours. \cite{Assef2018}, used the W1 and W2 criteria of \cite{Stern2012} and extended them to fainter magnitudes. However, infrared selection techniques are biased against low luminosity AGN. In a recent study, \cite{Pouliasis2020} showed that SED decomposition can provide a complementary tool to the X-ray and IR selection techniques, since it makes use of large wavelength ranges to efficiently disentangle accretion from star formation \citep{Ciesla2015, Dietrich2018, Malek2018}.

Our AGN sample consists of X-ray selected sources with high X-ray luminosity ($\rm {L_{X}>10^{42}\, erg\,s^{-1}}$), therefore the contamination from non AGN systems should be minimal. Thus, we wish to examine whether the addition of polar dust in the fitting process affects the efficiency of the SED decomposition to find a strong AGN component ($\rm frac_{AGN}$) and compare it with the efficiency of IR colour selection criteria to detect AGN candidates among X-ray sources. 

In Fig. \ref{assef_colour}, we plot the W1-W2 vs. W2 for the X-ray AGN in our sample. With magenta and grey, we mark the sources that are selected as IR AGN candidates with a confidence level of 75\% and 90\%, respectively, using the \cite{Assef2018} criteria. Circles present all the X-ray AGN in our dataset, detected by WISE, colour-coded based on the AGN fraction measurements of X-CIGALE. For estimating them, we have included the polar dust as a free parameter but their X-ray flux has been ignored. There are 1956 sources with WISE counterparts that also fulfill our selection criteria (section \ref{sec_finalsample}). 689/1956 ($\sim 35\%$) are detected by Assef et al. with 90\% confidence ($889/1956\approx 45\%$, with 75\% confidence). 671 out of the 689 sources ($\sim 97\%$), have $\rm frac_{AGN}>0.2$ ($631/689\approx 92\%$ with $\rm frac_{AGN}>0.3$), i.e. the SED decomposition reveals strong AGN contribution in the IR emission of the system. In section \ref{appendix_agnfrac}, we examine how many of the sources with measured $\rm frac_{AGN}>0.2$ (and $\rm frac_{AGN}>0.3$) have true $\rm frac_{AGN}<0.2$, i.e. the AGN fraction has been miscalculated and the AGN contribution in the IR emission of the galaxy is weaker than assumed. Our analysis shows $<10\%$ contamination from such systems (Fig. \ref{frac_mock_data2}). Among IR selected AGN with 75\% confidence, the numbers are $851/889\approx 96\%$ with $\rm frac_{AGN}>0.2$ ($795/889\approx 90\%$ with $\rm frac_{AGN}>0.3$). Therefore, X-CIGALE finds a strong AGN component in more than $90\%$ of the IR selected AGN. However, as we notice in Fig. \ref{assef_colour}, there is a large number of fainter AGN  that are missed by IR selection colour criteria, but the SED fitting finds strong AGN IR emission. Specifically, $85\%$ of the sources have $\rm frac_{AGN}>0.2$ and $74\%$ have $\rm frac_{AGN}>0.3$. Thus, even considering the more strict criterion of $\rm frac_{AGN}>0.3$, SED decomposition manages to detect a larger fraction of X-ray selected AGN than IR colour criteria ($74\%$ vs. $45\%$) while at the same time it also uncovers $>90\%$ of the IR colour selected AGN. Repeating the same exercise without the inclusion of polar dust in the SED fitting process yields, 71\% sources with $\rm frac_{AGN}>0.2$ and 55\% with $\rm frac_{AGN}>0.3$, among the non-IR detected AGN. Thus, the addition of polar dust allows SED decomposition to uncover a larger fraction of X-ray AGN. The results are summarised in Table \ref{table:detection_efficiency}. On the other hand, the addition of the X-ray flux only marginally improves the AGN detection efficiency. Specifically, among the non-IR selected AGN, modelling the SEDs with X-ray flux and polar dust, yields $87\%$ of the sources with $\rm frac_{AGN}>0.2$ and $75\%$ with $\rm frac_{AGN}>0.3$, compared to $85\%$ and $74\%$, respectively, using only the polar dust in the fitting process.

In Fig. \ref{donley_colour}, we repeat the same exercise using the $\it {Spitzer}$ colour selection criteria of \cite{Donley2012}. 968 AGN in our sample are detected by {\it Spitzer} and follow our selection criteria (section \ref{sec_finalsample}). 409/968 (42\%) are IR selected AGN, based on the criteria presented by \cite{Donley2012}. 98\% of these IR selected AGN have $\rm frac_{AGN}>0.2$ (92\% with $\rm frac_{AGN}>0.3$). Among the 968 sources with {\it{Spitzer}} counterparts, $\sim 80\%$ have $\rm frac_{AGN}>0.2$ and $\sim 67\%$ have $\rm frac_{AGN}>0.3$. These numbers drop to $\sim 65\%$ and $\sim 49\%$, when polar dust is not included in the SED fitting process. 

From this part of the analysis, we conclude that the addition of polar dust improves the ability of the algorithm to disentangle the AGN and host galaxy IR emission and thus increases the efficiency of the SED decomposition to detect a strong AGN component.

\section{Summary}
\label{sec_summary}

In this work, we use the SED fitting code X-CIGALE to model the SEDs of $\sim 2500$ X-ray AGN in the XMM-XXL field. X-CIGALE has some important features to model the AGN emission. It accounts for polar-dust extinction that is commonly found, especially in the case of X-ray selected AGN and includes X-ray data in the SED fitting process.  

Our analysis shows that the estimated $\rm L_{2500\,\AA}$ and $\rm L_{2\,keV}$ follow the adopted Just et al relation, as well as other similar relations in the literature \citep{Lusso2016}. The SED fitting results are not sensitive to the choice of the extinction law used and the temperature of polar dust. This result holds for our X-ray dataset and the available photometry. The addition of far-IR ({\it Herschel}) photometry does not statistically affect the polar dust and AGN fraction measurements, but, it slightly increases the statistical significance of the latter (from $3.6\sigma$ to $4.5\sigma$).  

About half of the sources identified as type 2 with the addition of X-ray flux, are found as type 1 in the absence of X-ray information. Visual inspection of randomly selected optical (SDSS) spectra, revealed that for $\sim85\%$ of the AGN that lie at $\rm z<1$, the optical spectrum agrees with the X-CIGALE's classification when using f$_X$. However, at higher redshifts ($z>1.5$) lack of available IR photometry does not allow the algorithm to classify sources in a robust manner. Inclusion of polar dust also improves the agreement of the AGN type classification between X-CIGALE and optical spectra. We conclude that the addition of the X-ray flux and polar dust in the fitting process, improves the accuracy of the classification of AGN into type 1 and 2. 

Compared to IR selection techniques \citep{Donley2012, Assef2018}, X-CIGALE recovered $>90\%$ of the IR selected AGN. Among the X-ray detected AGN that are not IR selected, SED decomposition attributes a large AGN component in the IR emission of the system ($\rm frac_{AGN}>0.2$) for $\sim 85\%$ of them. This number drops to $\sim 70\%$, when polar dust is not included in the SED modelling. Thus, addition of polar dust improves the efficiency of the SED decomposition to detect AGN.

One of the most important strengths of SED fitting is that it allows disentangling the IR emission of AGN from that of the host galaxy, quantified by $\rm frac_{AGN}$ in X-CIGALE. The new features of X-CIGALE improve the AGN fraction measurements. Specifically, the addition of the X-ray flux, improves the statistical significance of the AGN fraction measurements, in particular for luminous ($\rm {L_{X}>10^{45}\, erg\,s^{-1}}$) sources (Fig. \ref{frac_significance}). The addition of polar dust increases the $\rm frac_{AGN}$ estimations, since the AGN contribute more to the IR emission of the system. 

The conclusions of our analysis hold under the condition that (mid-) IR photometry is available in the SED fitting process. Lack of IR photometry may result in unreliable X-ray luminosity calculations, lower and less robust AGN fraction estimates and AGN type misclassifications.

\begin{acknowledgements}

\\
The authors thank the anonymous referee for their comments.
\\
The authors thank Raphael Shirley and Yannick Roehlly  for their help to retrieve IR fluxes on the XMM-LSS field.
\\
GM acknowledges support by the Agencia Estatal de Investigación, Unidad de Excelencia María de Maeztu, ref. MDM-2017-0765.
\\
MB acknowledges FONDECYT regular grant 1170618
\\
XXL  is  an  international  project  based  around  an  $\it{XMM}$  Very Large Programme surveying two 25 deg$^2$ extragalactic fields at a depth of $\sim$  6 $\times$ $10^{-15}$ erg cm $^{-2}$ s$^{-1}$ in the [0.5-2] keV band for point-like sources. The XXL website ishttp://irfu.cea.fr/xxl/.  Multi-band  information  and  spectroscopic  follow-up  of the X-ray sources are obtained through a number of survey programmes,  summarised  at http://xxlmultiwave.pbworks.com/.
\\
This research has made use of data obtained from the 3XMM XMM-\textit{Newton} 
serendipitous source catalogue compiled by the 10 institutes of the XMM-\textit{Newton} 
Survey Science Centre selected by ESA.
\\
This work is based on observations made with XMM-\textit{Newton}, an ESA science 
mission with instruments and contributions directly funded by ESA Member States 
and NASA. 
\\
Funding for the Sloan Digital Sky Survey IV has been provided by the Alfred P. Sloan Foundation, the U.S. Department of Energy Office of Science, and the Participating Institutions. SDSS-IV acknowledges
support and resources from the Center for High-Performance Computing at
the University of Utah. The SDSS web site is \url{www.sdss.org}.
\\
SDSS-IV is managed by the Astrophysical Research Consortium for the 
Participating Institutions of the SDSS Collaboration including the 
Brazilian Participation Group, the Carnegie Institution for Science, 
Carnegie Mellon University, the Chilean Participation Group, the French Participation Group, Harvard-Smithsonian Center for Astrophysics, 
Instituto de Astrof\'isica de Canarias, The Johns Hopkins University, 
Kavli Institute for the Physics and Mathematics of the Universe (IPMU) / 
University of Tokyo, Lawrence Berkeley National Laboratory, 
Leibniz Institut f\"ur Astrophysik Potsdam (AIP),  
Max-Planck-Institut f\"ur Astronomie (MPIA Heidelberg), 
Max-Planck-Institut f\"ur Astrophysik (MPA Garching), 
Max-Planck-Institut f\"ur Extraterrestrische Physik (MPE), 
National Astronomical Observatories of China, New Mexico State University, 
New York University, University of Notre Dame, 
Observat\'ario Nacional / MCTI, The Ohio State University, 
Pennsylvania State University, Shanghai Astronomical Observatory, 
United Kingdom Participation Group,
Universidad Nacional Aut\'onoma de M\'exico, University of Arizona, 
University of Colorado Boulder, University of Oxford, University of Portsmouth, 
University of Utah, University of Virginia, University of Washington, University of Wisconsin, 
Vanderbilt University, and Yale University.
\\
This publication makes use of data products from the Wide-field Infrared Survey 
Explorer, which is a joint project of the University of California, Los Angeles, 
and the Jet Propulsion Laboratory/California Institute of Technology, funded by 
the National Aeronautics and Space Administration. 
\\
The VISTA Data Flow System pipeline processing and science archive are described 
in \cite{Irwin2004}, \cite{Hambly2008} and \cite{Cross2012}. Based on 
observations obtained as part of the VISTA Hemisphere Survey, ESO Program, 
179.A-2010 (PI: McMahon). We have used data from the 3rd data release.
\\
This work is based [in part] on observations made with the Spitzer Space Telescope, which was operated by the Jet Propulsion Laboratory, California Institute of Technology under a contract with NASA
\\
The project has received funding from Excellence Initiative of Aix-Marseille University - AMIDEX, a French `Investissements d'Avenir' programme.

\end{acknowledgements}

\bibliography{mybib}{}
\bibliographystyle{aa}

\clearpage

\appendix

\section{Mock catalogue analysis}

As mentioned in section \ref{sec_mock}, X-CIGALE offers the option to create and analyse mock catalogues based on the best fit model of each source of the dataset. Here, we use this to assess the efficiency of X-CIGALE in estimating important parameters. Throughout this section, the bayesian estimates of the mock values will be presented.

\subsection{The efficiency of X-CIGALE in estimating $\alpha _{ox}$ and $\rm L_{2500\,\AA}$} 
\label{appendix_alpha_l2500}

In Fig. \ref{aox_mock_data}, we test X-CIGALE's reliability in estimating the $\alpha _{ox}$ parameter, using the results of the mock catalogue. Specifically, we compare the true values of the parameters from the best fit of the dataset to the Bayesian values of the parameter obtained from the fit of the mock catalogue. The distribution of the difference of the two values highly peaks at zero and $\sim 95\%$ of the sources are within $\pm 0.1$. This demonstrates that the algorithm can reliably recover this parameter. 

In the same fashion, Fig. \ref{l2500_mock_data}, presents the efficiency of X-CIGALE to constrain $\rm L_{2500\,\AA}$, with (left panel) and without (right panel) the X-ray flux. When the X-ray flux is included in the SED, $\rm L_{2500\,\AA}$ is well constrained. When there is no X-ray information in the SED the scatter is larger. This is expected, based on how X-CIGALE calculates this parameter in the absense of the X-ray flux (see section \ref{sec_agn_emission}).

\begin{figure}
\centering
  \includegraphics[width=1.\linewidth]{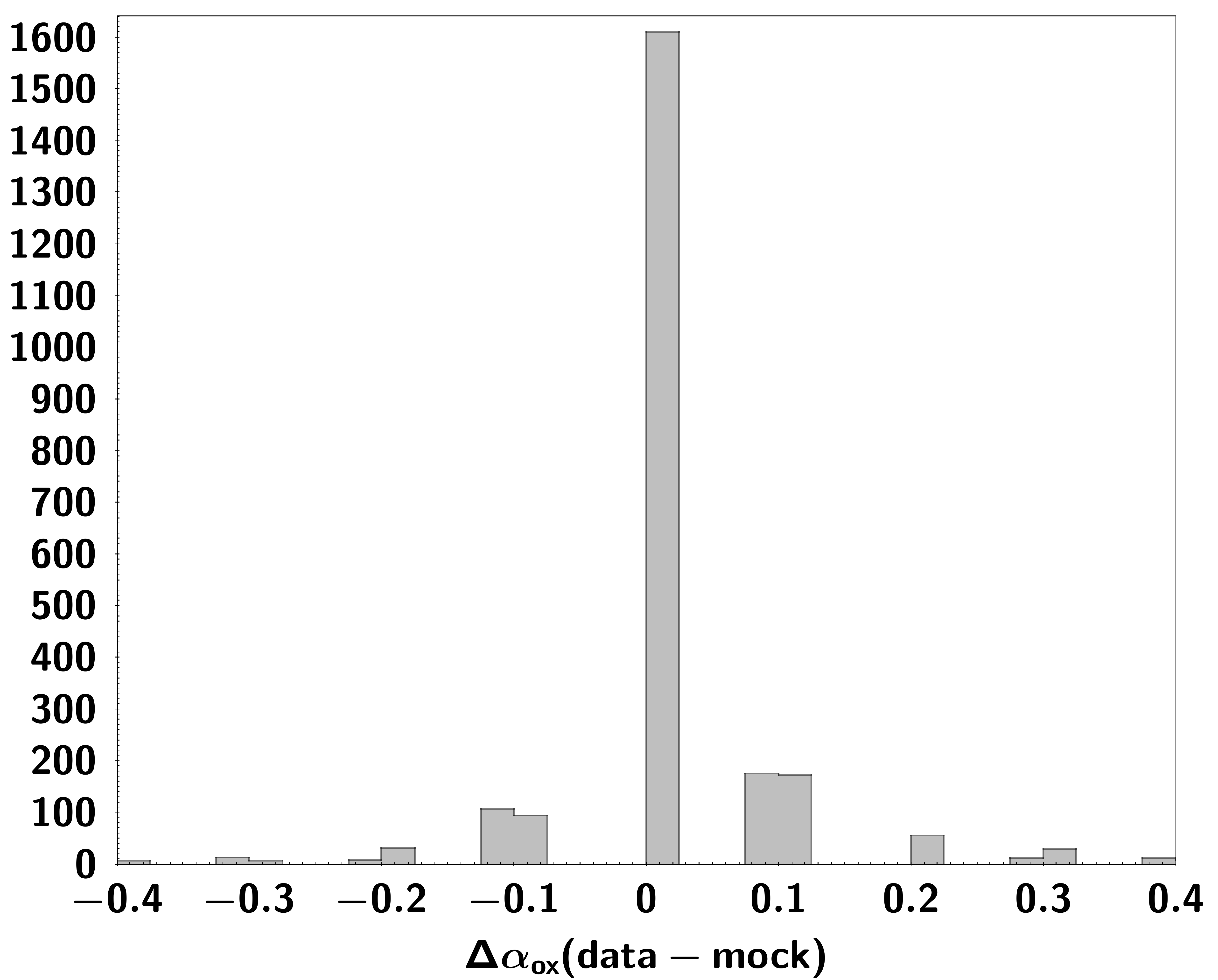}
  \caption{The efficiency of X-CIGALE to constrain the $\alpha _{ox}$ parameter, using the mock catalogue (see section \ref{sec_mock}). THe distribution highly peaks at zero, with $95\%$ of the sources to lie within $\pm 0.1$. This demonstrates that the algorithm can successfully estimate $\alpha _{ox}$.}
  \label{aox_mock_data}
\end{figure}

\begin{figure*}
\centering
\begin{subfigure}{.5\textwidth}
  \centering
  \includegraphics[width=1.\linewidth]{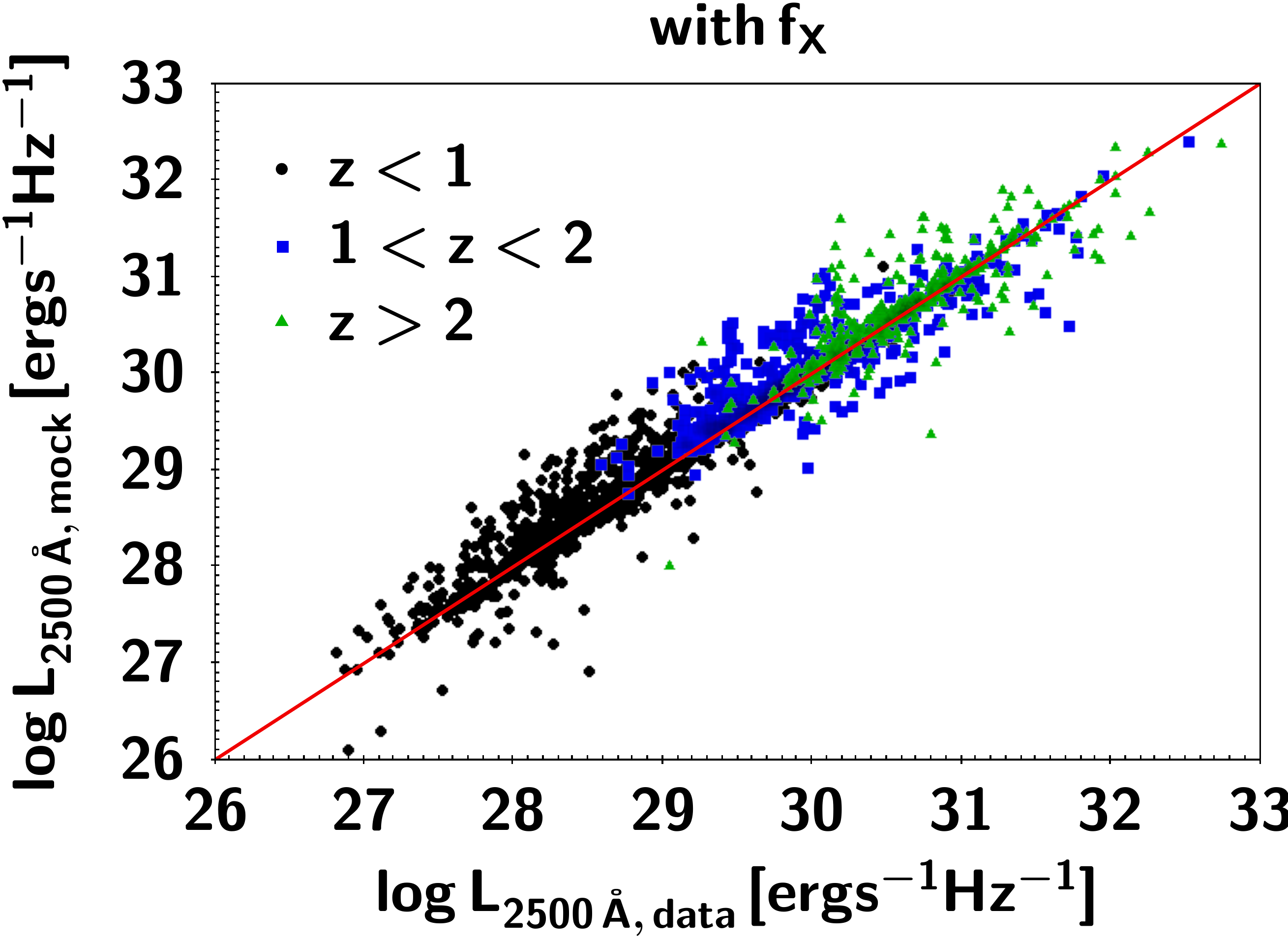}
  \label{l2500_mock_data_fxebv}
\end{subfigure}%
\begin{subfigure}{.5\textwidth}
  \centering
  \includegraphics[width=1.\linewidth]{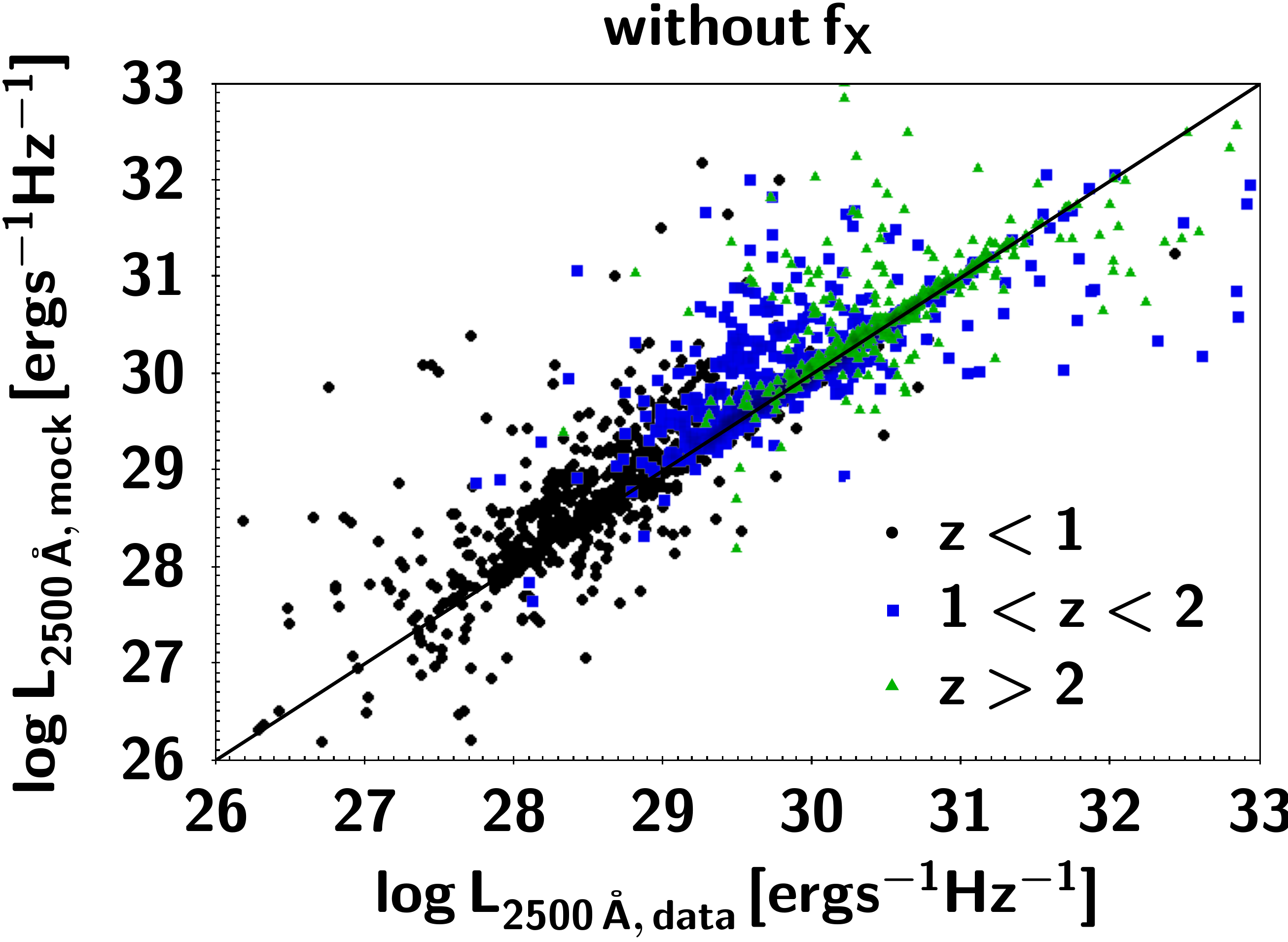}
  \label{l2500_mock_data_nofxebv}
\end{subfigure}
\caption{Comparison of the $\rm L_{2500\,\AA}$ values estimated by X-CIGALE for the mock sources with the true values (data). Left panel: When the X-ray flux is included in the SED, the $\rm L_{2500\,\AA}$ is well constained. Right panel: The X-ray flux is not included in the data. Although, X-CIGALE recovers the parameter successfully, the scatter is larger compared to the results with the X-ray flux.}
\label{l2500_mock_data}
\end{figure*}

\subsection{The efficiency of X-CIGALE in estimating the AGN fraction}
\label{appendix_agnfrac}

Here, we examine the accuracy of X-CIGALE in the estimation of the AGN fraction. In Fig. \ref{frac_mock_data2}, we plot the AGN fraction values estimated by the fit of the mock catalogue vs. the true values, i.e. the values from the best fit of the data. Circles present the mean measurements and their standard deviation is also plotted. Median values are shown by triangles. There is a good agreement between the estimated and the true values.

In Fig. \ref{frac_plot_mock_data_lx}, we plot the difference of the AGN fraction values estimated by fitting the mock catalogue from the true values. We present the results when the X-ray flux has been included in the SED (shaded histograms) and without including the X-ray flux (green lines). The results are split into X-ray luminosity bins. Based on our analysis, X-CIGALE can recover successfully the AGN fractions, regardless of whether the X-ray flux is included in the SED fitting or not, at all X-ray luminosities.

In section \ref{ebv_effect3}, we examine the fraction of X-ray AGN for which a strong AGN component is measured, i.e. $\rm frac_{AGN}>0.2$. In Fig. \ref{frac_mock_data3}, we test the level of contamination in these calculations, i.e., the percentage of sources that their true AGN fraction is $<0.2$, albeit the measured $\rm frac_{AGN}>0.2$. Using a threshold of measured (mock) $\rm frac_{AGN}>0.2$ the contamination is $\approx 10\%$ which drops to $\sim 6.5\%$ if we use a higher threshold of $\rm frac_{AGN}>0.3$. For this calculation we run X-CIGALE with polar dust and without X-ray flux, in accordance with the configuration used in section \ref{ebv_effect3}.

\begin{figure}
\centering
  \includegraphics[width=1.\linewidth]{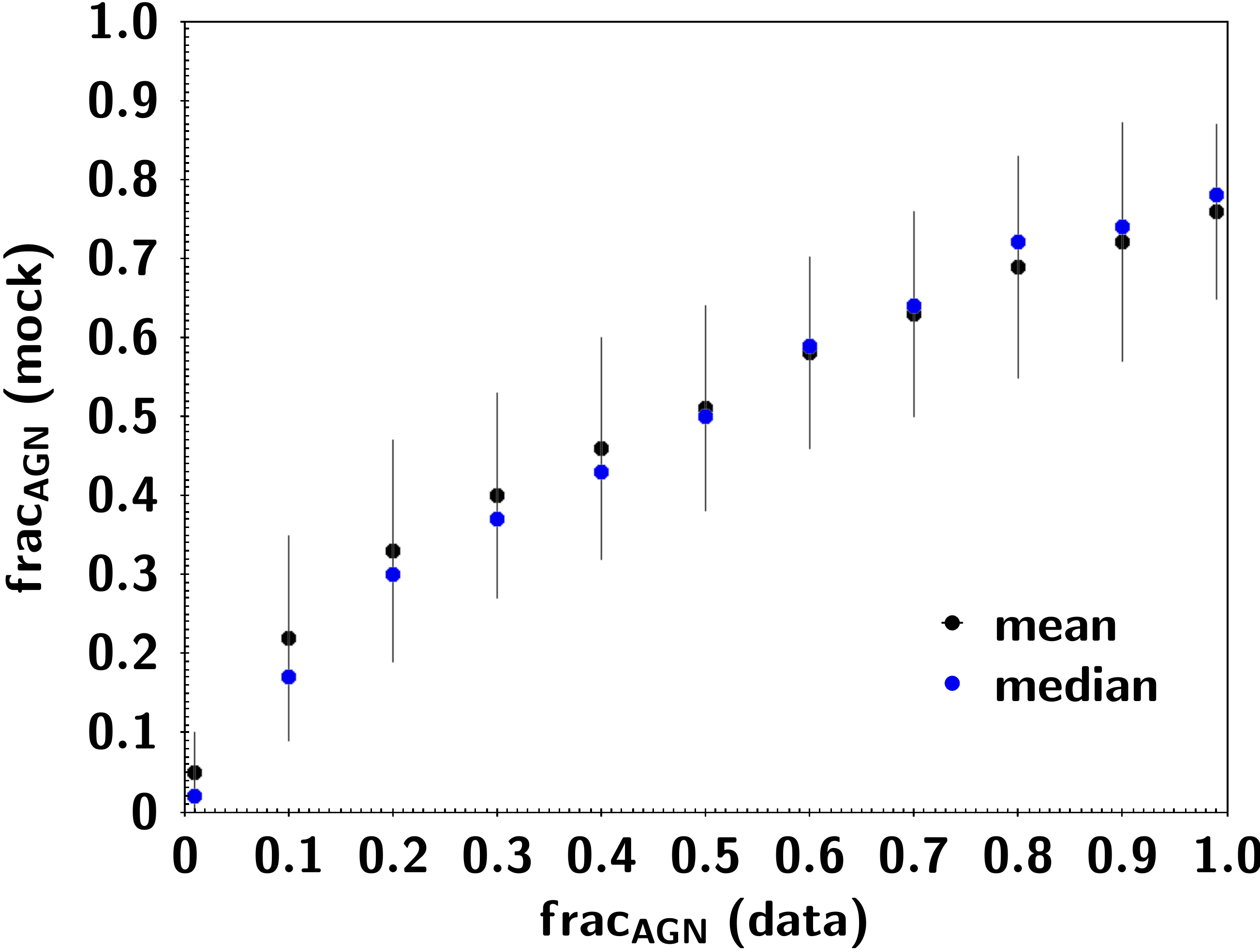}
  \caption{AGN fraction values estimated by the fit of the mock catalogue vs. the true values (from the fit of the data). Circles present the mean measurements and their standard deviation is also plotted. Median values are shown by triangles.}
  \label{frac_mock_data2}
\end{figure}

\begin{figure}
\centering
\begin{subfigure}{.335\textwidth}
  \centering
  \includegraphics[width=1.\linewidth]{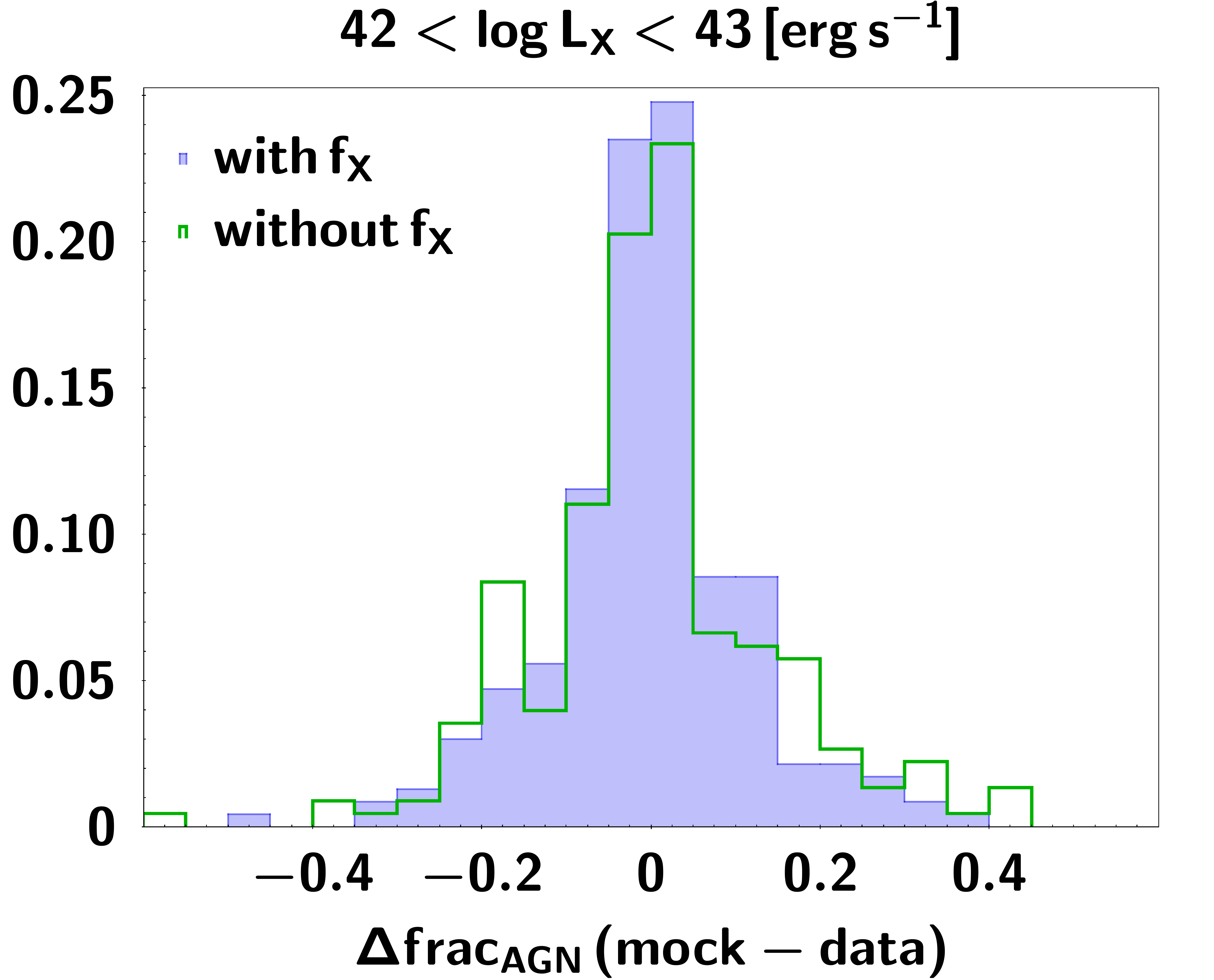}
  \label{frac_mock_data_fxebv}
\end{subfigure}
\begin{subfigure}{.335\textwidth}
  \centering
  \includegraphics[width=1.\linewidth]{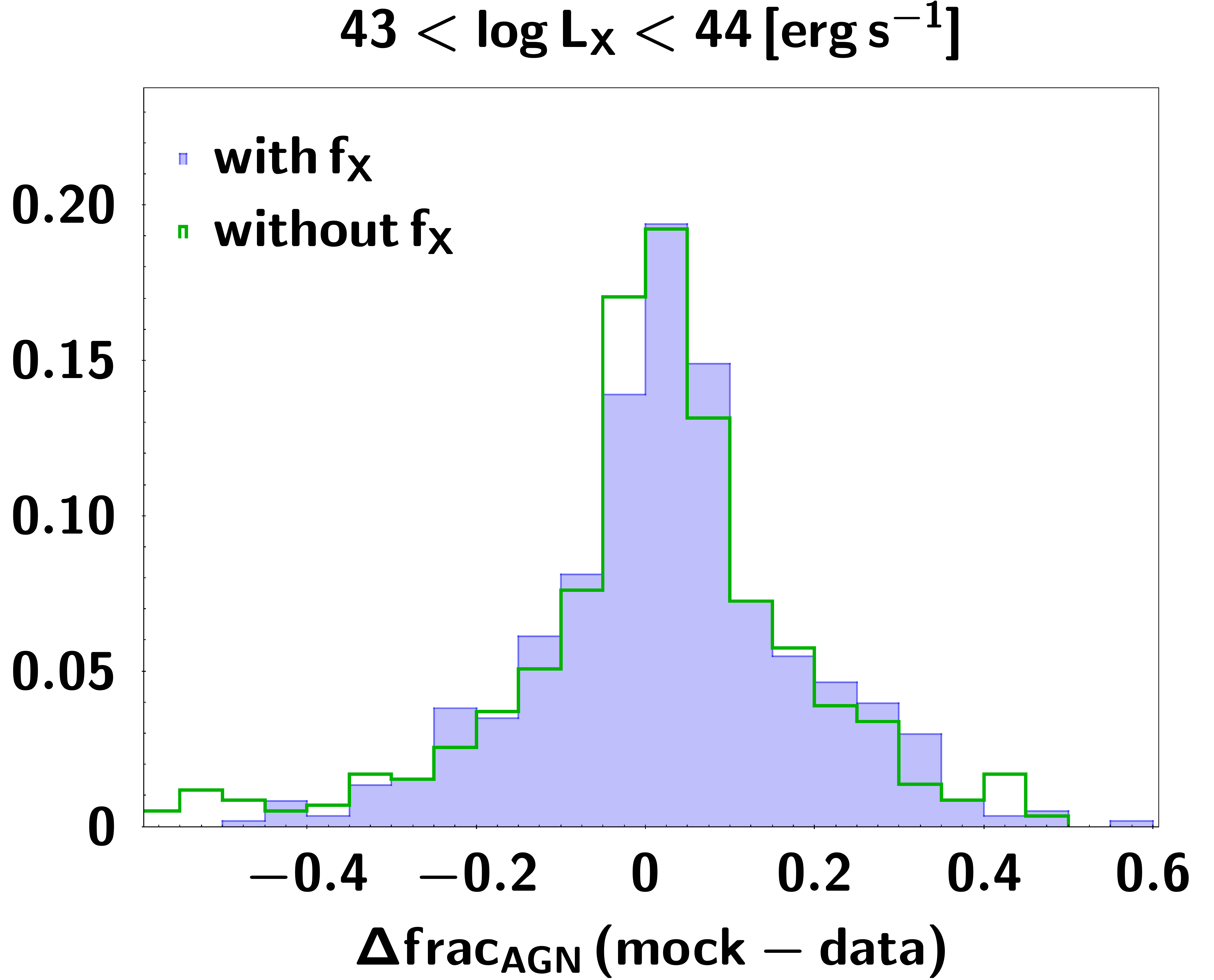}
  \label{frac_mock_data_nofxebv}
\end{subfigure}
\begin{subfigure}{.335\textwidth}
  \centering
  \includegraphics[width=1.\linewidth]{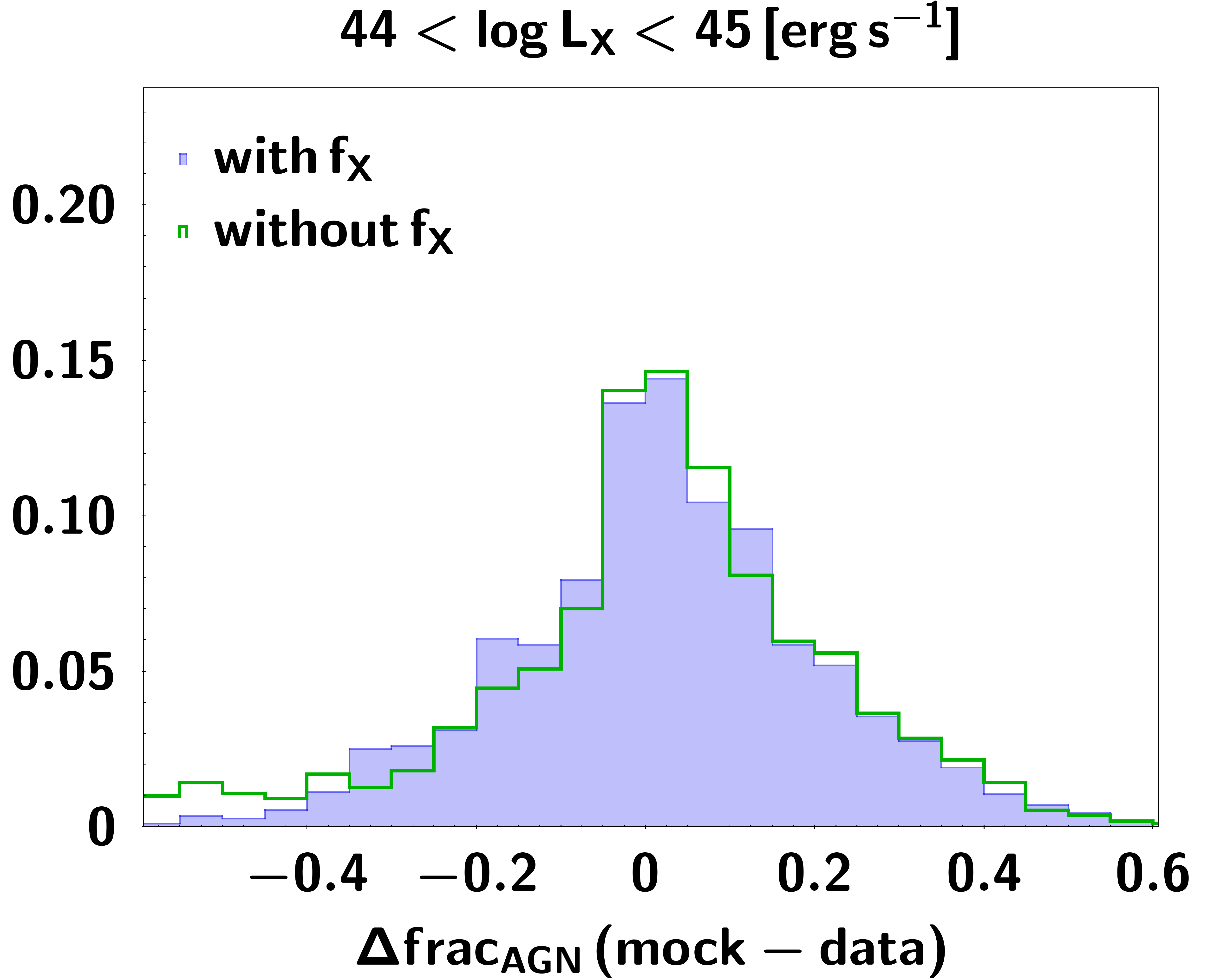}
  \label{frac_mock_data_nofxebv}
\end{subfigure}
\begin{subfigure}{.335\textwidth}
  \centering
  \includegraphics[width=1.\linewidth]{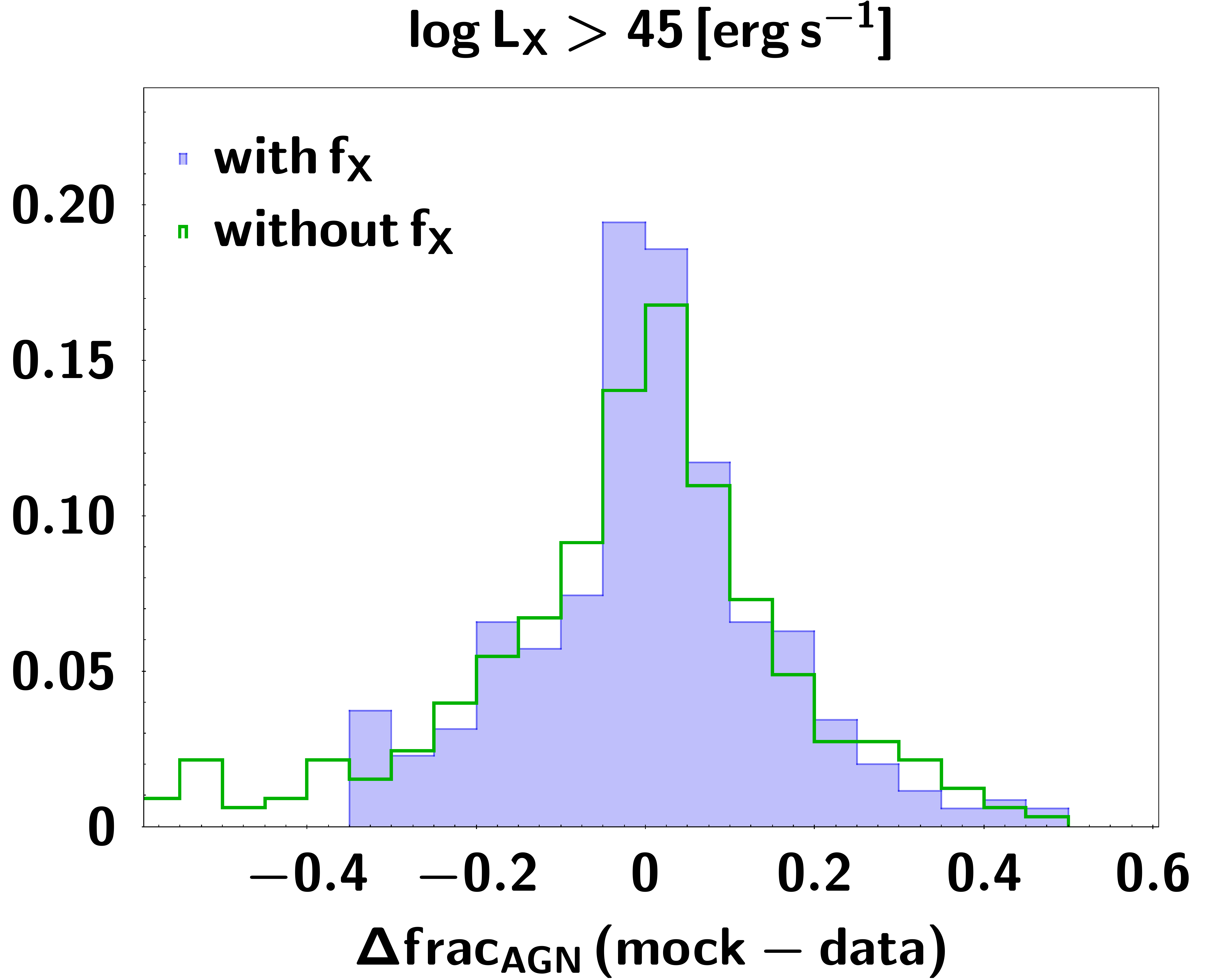}
  \label{frac_mock_data_nofxebv}
\end{subfigure}
\caption{The difference of the AGN fraction values estimated by fitting the mock catalogue from the true values. We present the results when the X-ray flux has been included in the SED (shaded histograms) and without including the X-ray flux (green histograms). The results are split into X-ray luminosity bins. X-CIGALE can recover successfully the AGN fractions, regardless of whether the X-ray flux is included in the SED fitting or not, at all X-ray luminosities.}
\label{frac_plot_mock_data_lx}
\end{figure}

\begin{figure}
\centering
  \includegraphics[width=1.\linewidth]{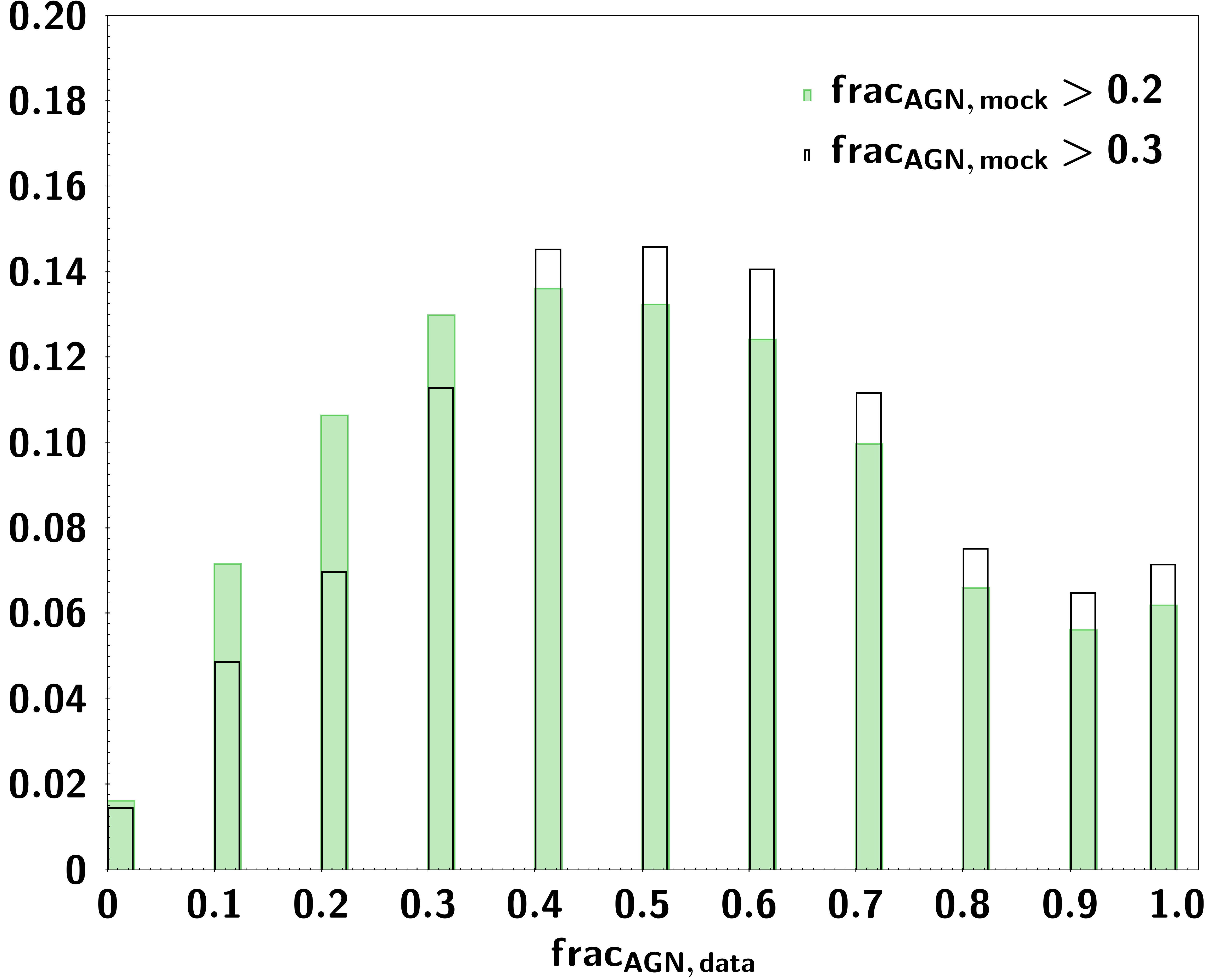}
  \caption{Binned measurements of the true AGN fraction values (data), for $\rm frac_{AGN, mock}>0.2$ (green histogram) and $\rm frac_{AGN, mock}>0.3$ (black histogram) from the mock analysis. For sources with $\rm frac_{AGN, mock}>0.2$, there is $\approx 10\%$ contamination, i.e. sources that the AGN component is low  $\rm frac_{AGN, data}<0.2$. For sources with $\rm frac_{AGN, mock}>0.3$, the contamination is $\sim 6.5\%$.}
  \label{frac_mock_data3}
\end{figure}

\subsection{The efficiency of X-CIGALE in the estimation of polar dust contribution}
\label{appendix_polar}

To check whether X-CIGALE can successfully constrain the effect of dust extinction by polar dust, in Fig. \ref{ebv_data_mock} we plot the Bayesian estimations of the $\rm E_ {B-V}$ parameter for the polar dust vs. its exact value. The scatter (1\,$\sigma$ variations) is substantial. Furthermore, for $\rm E_ {B-V}<0.5$ the fit is not sensitive to an incremental increase of the reddening parameter. The sensitivity is better at higher values of $\rm E_ {B-V}$, but the parameter is still not determined well. The picture remains the same regardless of whether the X-ray flux is taken into account in the SED fitting. These results suggest that it is redundant to use in the SED configuration process, values of $\rm E_ {B-V}$ within small intervals and that the estimated values of the parameter are not well constrained.

\begin{figure*}
\centering
\begin{subfigure}{.5\textwidth}
  \centering
  \includegraphics[width=1.\linewidth]{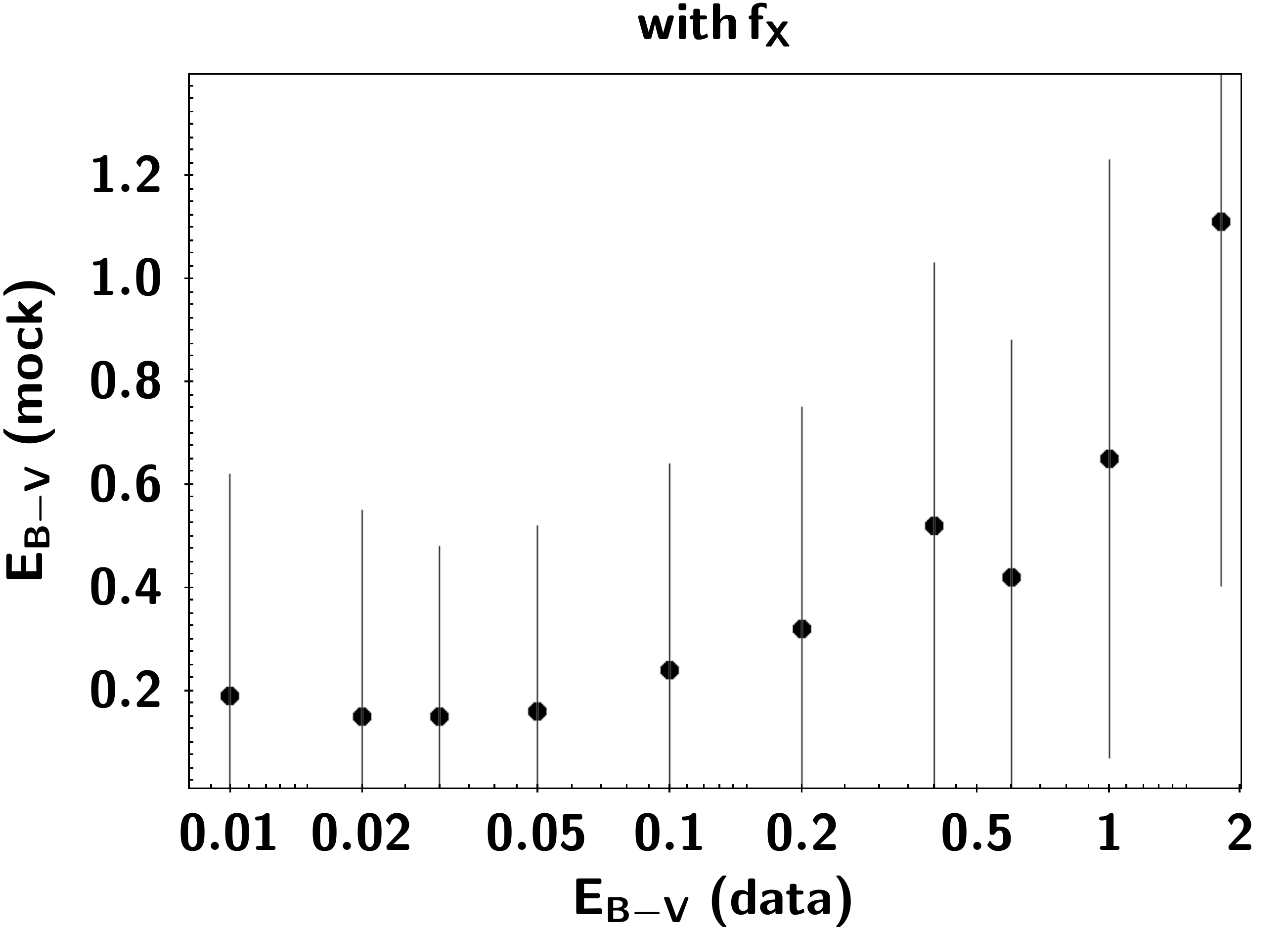}
  \label{ebv_data_mock_fxebv}
\end{subfigure}%
\begin{subfigure}{.5\textwidth}
  \centering
  \includegraphics[width=1.\linewidth]{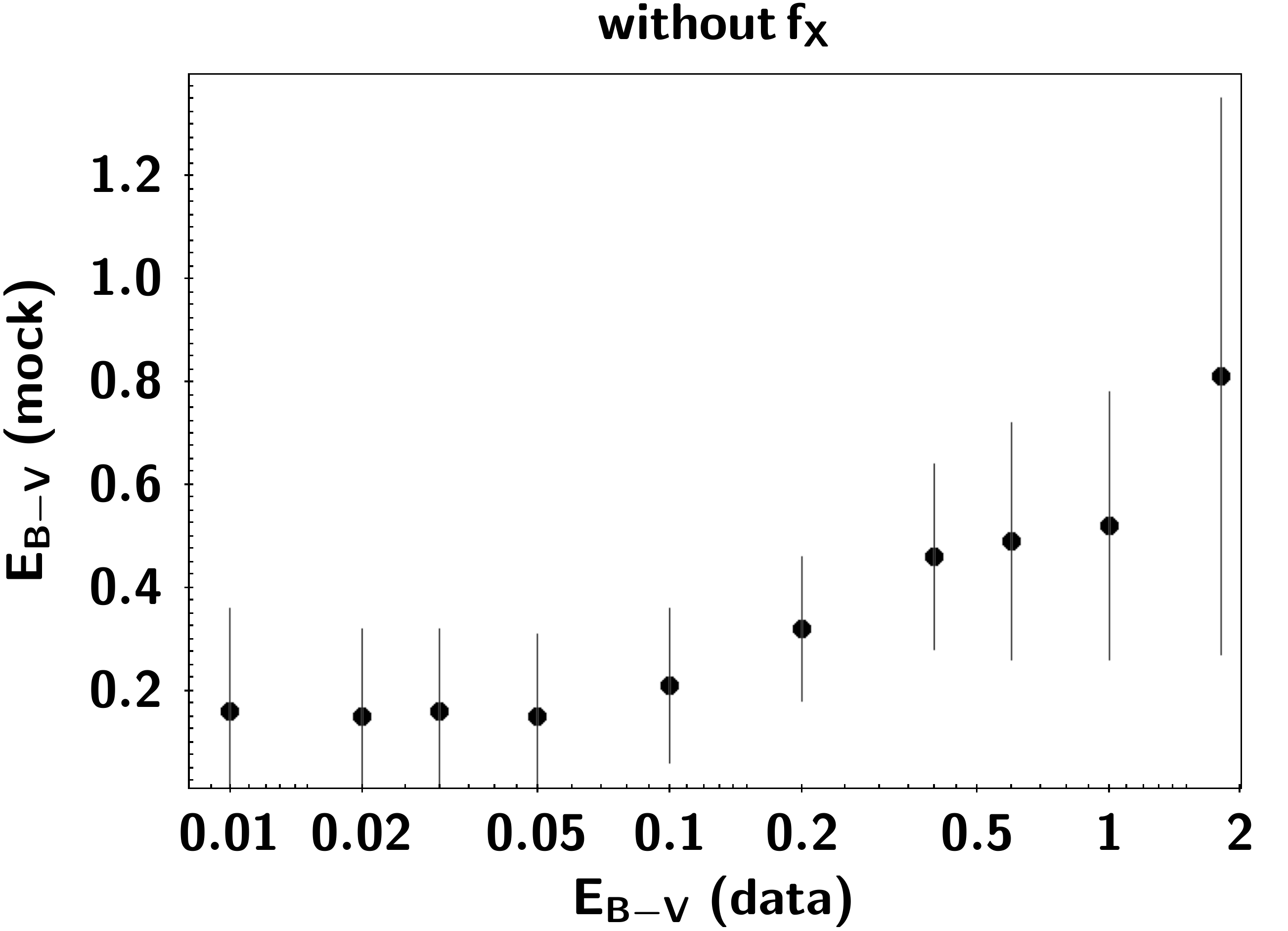}
  \label{ebv_data_mock_nofxebv}
\end{subfigure}
\caption{Polar dust estimations from the mock sources vs. the input (data) values. Although the 1\,$\sigma$ deviation is large, mock values increase with the increase of the data values. However, the algorithm is not sensitive to incremental increases of the parameter. The trends are similar regardless, of whether the X-ray flux is included (left panel) or not (right panel) in the fitting process.}
\label{ebv_data_mock}
\end{figure*}

\end{document}